\documentclass[reqno,11pt]{amsart}
\usepackage{hyperref}
\usepackage[mathscr]{eucal}
\usepackage{enumerate}
\numberwithin{equation}{section}

\newtheorem{Def}{Definition}[section]
\newtheorem{Thm}[Def]{Theorem}
\newtheorem{Prop}[Def]{Proposition}
\newtheorem{Lemma}[Def]{Lemma}

\newcommand{\beq}{\begin{equation}}
\newcommand{\eeq}{\end{equation}}
\newcommand{\Proof}{\begin{proof}}
\newcommand{\QED}{\end{proof} \noindent}

\newcommand{\Riem}{{\rm Riem}}
\newcommand{\mm}{\hspace{-.08cm}\cdot \hspace{-.08cm}}

\newcommand{\R}{\mathbb{R}}

\newcommand{\Gammati}{\tilde{\Gamma}}

\DeclareMathOperator{\tr}{tr}

\allowdisplaybreaks[1]

\title[Regularity Singularities]{The Regularity Transformation Equations:\\ An elliptic mechanism for smoothing gravitational metrics in General Relativity}

\author[M.\ Reintjes]{Moritz Reintjes}
\address{Fachbereich f\"ur Mathematik und Statistik \\ Universit\"at Konstanz \\ D-78467 \\ Germany}
\email{moritzreintjes@gmail.com}
\thanks{M. Reintjes is currently supported by the German Research Foundation, DFG grant FR822/10-1, and was supported by FCT/Portugal through (GPSEinstein) PTDC/MAT-ANA/1275/2014 and UID/MAT/04459/2013 from January 2017 until December 2018.}

\author[B.\ Temple]{Blake Temple \\ \\ May 29, 2018 }
\address{Department of Mathematics\\ University of California\\ Davis, CA 95616\\ USA}
 \email{temple@math.ucdavis.edu}

\begin{document}

\begin{abstract}
A central question in General Relativity (GR) is how to determine whether singularities are geometrical properties of spacetime, or simply anomalies of a coordinate system used to parameterize the spacetime.   In particular, it is an open problem whether there always exist coordinate transformations which smooth a gravitational metric to optimal regularity, two full derivatives above the curvature tensor, or whether {\it regularity singularities} exist.  We resolve this open problem above a threshold level of smoothness by proving in this paper that the existence of such coordinate transformations is equivalent to solving a system of nonlinear elliptic equations in the unknown Jacobian and transformed connection, both viewed as matrix valued differential forms. We name these the {\it Regularity Transformation equations}, or {\it RT-equations}.  In a companion paper we prove existence of solutions to the RT-equations for connections $\Gamma\in W^{m,p},$ curvature ${\rm Riem}(\Gamma) \in W^{m,p}$, assuming $m\geq1$, $p>n$. Taken together, these results imply that there always exist coordinate transformations which smooth arbitrary connections to optimal regularity, (one derivative more regular than the curvature), and there are no {\it regularity singularities}, above the threshold $m\geq1$, $p>n$.    Authors are currently working on extending these methods to the case of GR shock waves, when gravitational metrics are only Lipschitz continuous, ($m=0$, $p=\infty$), and optimal regularity is required to recover basic properties of spacetime. 
\end{abstract}

\maketitle

\section{Introduction}

Solutions of the Einstein equations of General Relativity (GR) are constructed in coordinate systems in which the associated partial differential equations (PDE's) take on a solvable form.  A very first question in GR is then, which properties of the spacetime represent the true geometry, and which are merely anomalies of the particular coordinate system used to parameterize the spacetime?  Here we consider solutions of the Einstein equations which appear to be {\it singular} in the sense that the metric connection fails to be one derivative smoother than the curvature tensor in the coordinate system in which the solutions are constructed.   For any such non-optimal spacetime, we prove the existence of coordinate transformations which smooth the metric and connection to optimal regularity, is equivalent to solving a system of nonlinear elliptic equations which we name the {\it RT-equations}, (for {\it Regularity Transformation equations}), equations in the unknown Jacobian and transformed connection, viewed as matrix valued differential forms.   This, together with an existence theory for the RT-equations presented in the companion paper \cite{ReintjesTemple_ell2}, establishes that there always exist coordinate transformations which smooth connections of arbitrary signature to optimal regularity, (one derivative more regular than the curvature). Thus there are no {\it regularity singularities}, for connections $\Gamma\in W^{m,p}$ with $Riem(\Gamma)\in W^{m,p}$, assuming $m\geq1$, $p>n$.  In words, a crinkled map of spacetime can always be smoothed by coordinate transformation above a threshold level of regularity.\footnote{The framework introduced in this paper and the existence theory in \cite{ReintjesTemple_ell2} are the basis for proving optimal regularity and Uhlenbeck compactness for $L^\infty$ connections in our followup paper \cite{ReintjesTemple_ell4}. The results in \cite{ReintjesTemple_ell2,ReintjesTemple_ell4} and in this paper are summarized in our RSPA article \cite{ReintjesTemple_ell3}.}    

To derive the RT-equations we develop an Euclidian Cartan algebra of matrix valued differential forms, and introduce special operations which have no scalar analogue.  Remarkable to us is that the RT-equations reduce the question of regularity singularities in {\it Lorentzian} spacetimes, to an existence problem for a system of {\it elliptic} Poisson-type equations.  As a corollary we see that metric signature is of no relevance to the question of regularity singularities.  In the lower regularity case of shock waves, e.g., $\Gamma,Riem(\Gamma)\in L^{\infty}$, the RT-equations place the problem of regularity singularities within the well-studied framework of elliptic regularity theory, connecting them to classical Calder\'{o}n-Zygmund singularities in elliptic PDE theory,\footnote{By this we mean counterexamples demonstrating that solutions of the linear Poisson equation are not always in $C^{1,1}$ when the sources are in $L^{\infty}$, \cite{internetnote}.} but the existence theory is more problematic.\footnote{This is addressed in authors' subsequent research \cite{ReintjesTemple_ell4}.} 

The problem of optimal regularity for Riemannian metrics with curvature tensors of low regularity was first addressed by Kazdan and DeTurck \cite{DeTurckKazdan}. In the case of Lorentzian metrics, this problem was first addressed by Anderson in \cite{Anderson} for GR vacuum spacetimes, and for non-vacuum spacetimes, subject to certain assumptions on the regularity of the Einstein tensor on the foliation assumed.  Anderson's results were revisited in \cite{ChenLeFloch} in the case of GR vacuum solutions.  Results on $L^2$-curvature-solutions of the vacuum Einstein equations in \cite{Klainermann} do not establish optimal regularity.   The above papers do not address GR shock waves, the case when the matter sources are non-zero, \cite{Israel,SmollerTemple,ReintjesTemple1}.    Our new approach to optimal metric regularity based on the RT-equations is different from these earlier approaches in that it does not employ any {\it apriori} coordinate ansatz, like wave gauge or harmonic coordinates, etc.  This eliminates the need for any additional assumptions beyond what is required to formulate the question of optimal regularity.  Our view at the start was that the coordinate systems of optimal regularity are, in general, too difficult to guess apriori, and one should seek equations for the optimal coordinates themselves.  These are realized in the RT-equations.   
 
GR shock waves provide an intriguing motivating example of non-optimal solutions of the Einstein equations, \cite{Israel,SmollerTemple,GroahTemple,ReintjesTemple1,VoglerTemple,LeFlochStewart}.  In \cite{GroahTemple}, shock wave solutions of the Einstein equations generated by the Glimm scheme could only be constructed in coordinate systems in which the metric is only Lipschitz continuous ($C^{0,1}$) at shocks, even though both the connection and curvature tensor of such solutions stay bounded in $L^{\infty}$. That is,  the connection is non-optimal in the sense that it is no more regular than the Riemann curvature tensor.   The question as to whether such $C^{0,1}$ metrics can always be smoothed one order to optimal regularity $C^{1,1}$ by coordinate transformation, is intimately related to the existence of locally inertial coordinate systems, the Hawking-Penrose singularity theorems, the equivalence of strong and weak solutions of the Einstein equations, and the local correspondence of GR with the physics of Special Relativity, \cite{GrafGrant,HawkingEllis,ReintjesTemple1}.  In the RSPA publication \cite{ReintjesTemple1} authors began by making the point that if such coordinate systems do not always exist, then non-optimality of the metric would be geometric, and hence shock waves would create new kinds of mild singularities which the authors termed {\it regularity singularities},  (see also \cite{Reintjes,ReintjesTemple_wave}).  We start here by generalizing the notion ``regularity singularity'' to arbitrary connections by defining it to be any point where the connection is non-optimal in the sense that it fails to be one full derivative more regular than its curvature tensor, in any coordinate system within the atlas of coordinate transformations whose Jacobians are one level more regular than the connection.   We state this precisely in the following definition:

\begin{Def}   \label{Def_regularity_singularity}
Let $\Gamma$ be a connection given in a coordinate system $x$ such that (each component of) its Riemann curvature tensor $Riem(\Gamma)$ is in $W^{m,p}$ for some $m\geq 0$, $p\geq 1$, but is no smoother in the sense that $Riem(\Gamma)$ is not in $W^{m',p}$ for any $m'>m$. We say $\Gamma$ has {\rm optimal regularity} in $x$-coordinates if $\Gamma \in W^{m+1,p}$, (one order smoother than $Riem(\Gamma)$).  We say $\Gamma$ has a  {\rm regularity singularity} at a point $q$ if $\Gamma$ fails to transform to optimal regularity under any $W^{m+2,p}$ coordinate transformation $x\to y$ defined in a neighborhood of $q$.\footnote{We recover the notion of regularity singularity for GR shock waves when $p=\infty$, $m=0$, c.f. \cite{ReintjesTemple_geo}.}  
\end{Def}

To motivate the definition, and to understand the role non-optimal solutions of the Einstein equations play in GR, consider the following: if one were to construct a solution to the Einstein equations $G=\kappa T$ in a given coordinate system $y$ in which the equations produce a unique optimal solution within a given smoothness class, say metric $g\in W^{m+2,p}$, connection $\Gamma\in W^{m+1,p}$,  and $Riem(\Gamma)\in W^{m,p}$, then application of a transformation $y\to x$ with Jacobian $J\in W^{m+1,p}$ will in general lower the regularity of the metric and its connection $\Gamma$ by one order, but will preserve $Riem(\Gamma)\in W^{m,p}$ because it is a tensor.   Thus, if that existence theory were posed in $x$-coordinates, it would produce the unique transformed solution $g\in W^{m+1,p}$, $\Gamma\in W^{m,p}$, and $Riem(\Gamma)\in W^{m,p}$.   In the latter case, we would not know that our unique solution exhibited optimal smoothness in a different coordinate system without knowing about the existence of the inverse transformation $x\to y$.   In this paper we address  the problem of reversing this argument, i.e., given a non-optimal metric in $x$-coordinates, does a smoothing transformation $x\to y$ always exist, and if so, how do we find it? 

To illustrate the difficulty in the problem of raising the metric regularity by coordinate transformation, consider the case of shock waves, where the components of the gravitational metric tensor $g_{ij}$ are $C^{0,1}$ functions, and the associated connection coefficients $\Gamma^i_{ljk}$ and Riemann curvature tensor $R^i_{jkl}$ are $L^{\infty}$ functions given in a fixed coordinate system $x$.   Since the metric transforms as a tensor, for such a transformation $x\to y$ to exist such that the components\footnote{We use the Einstein summation convention assuming summation over repeated up-down indices, and let Latin indices to denote $x$-coordinates and Greek indices to denote $y$-coordinates.} $g_{\alpha\beta}=J^i_{\alpha}g_{ij}J^j_{\beta}$ in $y$-coordinates are $C^{1,1}$,  would require that discontinuities in derivatives of $J$ cancel the discontinuities in derivatives of $g_{ij}$ in the Leibniz products that arise from taking derivatives of the transformed metric, to make 
\begin{eqnarray} \label{transform}
\frac{\partial}{\partial y^{\alpha}}g_{\alpha\beta}=\frac{\partial}{\partial y^{\alpha}}\left\{J^i_{\alpha}g_{ij}J^j_{\beta}\right\}\in C^{0,1}.
\end{eqnarray}  
This requires the Jacobian $J$ to have the same regularity as the metric. Thus the regularity of the mappings $x\to y$ and the atlas these generate should be $C^{1,1}$, one order smoother than the metric.  This atlas then preserves the $L^{\infty}$ regularity of $Riem(\Gamma)$ because the curvature transforms as a tensor.  Condition (\ref{transform}) on $J$ is in general impossible to meet for arbitrary Lipschitz metrics, for example, whenever the metric curvature contains delta function sources, \cite{ReintjesTemple1,SmollerTemple}.   Thus, the existence of such a smoothing transformation requires the assumption that the Riemann curvature tensor be in $L^{\infty}$, an additional constraint because the curvature involves second derivatives of $g$.   To enlighten the point, note that a metric $g$ is always one order smoother than its connection $\Gamma$ by Christoffel's formulas, however $\Gamma$ need not be one derivative smoother than $Riem(\Gamma)$ because the curvature only involves the exterior derivatives $d\Gamma$, not the co-derivatives $\delta\Gamma$, (see Section \ref{Sec_Prelim} below).  These difficulties in smoothing metrics in $C^{0,1}$ carry over essentially unchanged to the problem of smoothing non-optimal connections at all levels of regularity, $\Gamma\in W^{m,p}$.   Our argument based on the RT-equations here establishes that, for $m\geq1$, $p>n$, the {\it only} assumption required to lift the regularity of a connection one order by coordinate transformation, is the assumption that $d\Gamma$, and hence the curvature, have the same regularity as the connection, and no additional assumptions on the regularity of the co-derivatives $\delta\Gamma$ are required.   

The authors' research program regarding the smoothing of non-optimal metrics in GR arose from the study of GR shock waves.    At smooth non-interacting shock surfaces, coordinate transformation to Gaussian normal coordinates at the surface, suffices to smooth an $L^{\infty}$ gravitational connection by one order to $C^{0,1}$ at shocks, by a now classical result of Israel in 1966 \cite{Israel}.\footnote{The Riemann normal coordinate construction for locally inertial frames is problematic for metrics $g\in C^{0,1}$, and is not sufficient to smooth a metric or a connection in general. } But for more general shock wave interactions, the only result we have is due to Reintjes \cite{Reintjes}, who proved that the gravitational metric can always be smoothed one order to $C^{1,1}$ in a neighborhood of the interaction of two shock waves from different characteristic families, in spherically symmetric spacetimes. Reintjes' procedure for finding the local coordinate systems of optimal smoothness is orders of magnitude more complicated than the Riemann normal, or Gaussian normal construction process.   The coordinate systems of optimal $C^{1,1}$ regularity are constructed in \cite{Reintjes} by solving a complicated non-local PDE highly tuned to the structure of the interaction. Trying to guess the coordinate system of optimal smoothness {\it apriori}, for example harmonic or Gaussian normal coordinates \cite{Choquet}, didn't work.  In Reintjes' construction, several apparent miracles happen in which the Rankine-Hugoniot jump conditions come in to make seemingly over-determined equations consistent, but, the principle behind what PDE's must be solved to smooth the metric in general, or when this is possible, appears entirely mysterious.  Extending Reintjes' result to general GR shock waves remains an open problem.    

Our first general principle regarding metric smoothing which did not rely on any specific structure of the shock interaction surfaces, came with the authors' discovery of the {\it Riemann-flat condition}, a necessary and sufficient condition for the existence of a coordinate transformation which smooths a connection in $L^\infty$ to $C^{0,1}$, \cite{ReintjesTemple_geo}. The Riemann-flat condition is the condition that there should exist a tensor $\Gammati$, one order smoother than $\Gamma$, such that  ${\rm Riem}(\Gamma-\Gammati)=0$. Now $\Gammati$ is continuous, so $\Gamma$ and $\Gamma -\Gammati$ have the same singular set of shock discontinuities.  Thus at first we thought the Riemann-flat condition was telling us that to smooth an $L^\infty$ shock wave connection one needed to extend the singular shock set to a flat connection by some sort of Nash embedding theorem. Our point of view changed with the successful idea that we might use the Riemann-flat condition to derive a system of elliptic equations, equivalent to the Riemann-flat condition. This led to the discovery of RT-equations, and we prove here that the existence of solutions to these equations is equivalent to the Riemann-flat condition. 

In this paper we establish the theory of metric smoothing based on the RT-equations for $\Gamma, d\Gamma\in W^{m,p}$,  $m\geq1$, $p>n$, so ${\rm Riem}(\Gamma)\in W^{m,p}$, and for metrics $g\in W^{m+1,p}$. By this we mean that component functions $\Gamma^i_{jk}(x)$ are in $W^{m,p}$ in some given coordinate system $x$, and $d\Gamma$ denotes the exterior derivative of $\Gamma$ viewed as a matrix valued $1$-form.  The case $m\geq1$, $p>n$ casts the theory of the RT-equations in its cleanest form, but assumes one order of regularity above the shock wave case $m=0$, $p=\infty$.  The Riemann-flat condition extends easily to higher regularities.   By Gaffney's inequality, our assumption $\Gamma,d\Gamma\in W^{m,p}$ implies all of the loss of derivative in $\Gamma$ occurs in $\delta\Gamma$, c.f. \cite{Dac} and \eqref{Gaffney} below.  Here we derive the RT-equations and prove Theorem \ref{Thm_main} which gives the equivalence of the RT-equations with the Riemann-flat condition when $m\geq1$, $p>n$.  In \cite{ReintjesTemple_ell2} we give the existence theory for the RT-equations when $m\geq1$, $p>n.$  Taken together, these results prove that there always exists a coordinate transformation $x\to y$ with Jacobian  $J\in W^{m+1,p}$, such that in $y$-coordinates, the connection is one degree smoother, $\Gamma\in W^{m+1,p}$, so long as $m\geq 1$, $p>n$.   Authors are currently working on extending these methods to the lower regularity of GR shock waves.\footnote{Since the writing of this paper, this result was established in \cite{ReintjesTemple_ell4}.}

\section{Statement of results} \label{Sec_results}

To state the main theorem, assume $\Gamma$ is a connection given by components in a coordinate system $x$, and view $\Gamma\equiv\Gamma^\mu_{\nu k}dx^k$ as a matrix valued $1$-form. The unknowns in the RT-equations are $\Gammati, J, A$ also taken to be matrix valued differential forms as follows: $J\equiv J^\mu_\nu$ is the Jacobian of the sought after coordinate transformation which smooths the connection, viewed as a matrix-valued $0$-form; $\Gammati\equiv \Gammati^\mu_{\nu k}dx^k$ is the unknown tensor one order smoother than $\Gamma$ such that $Riem(\Gamma-\Gammati)=0$, viewed as a matrix-valued $1$-form; and $A\equiv A^\mu_\nu$ is an auxiliary matrix valued $0$-form introduced to impose $Curl( J)=0$, the integrability condition for the Jacobian.

\begin{Thm} \label{Thm_main}
Assume $\Gamma$ is defined in a fixed coordinate system $x$ on $\Omega$, for $\Omega\subset{\mathbb R}^n$ open and with smooth boundary.   Assume that $\Gamma\in W^{m,p}(\Omega)$ and $d\Gamma\in W^{m,p}(\Omega)$ for $m\geq 1,$ $p>n$. Then the following equivalence holds: \vspace{.15cm} \newline 
Assume there exists $J \in W^{m+1,p}(\Omega)$ invertible, $\Gammati \in W^{m+1,p}(\Omega)$  and $A\in W^{m,p}(\Omega)$ which solve the elliptic system
\begin{eqnarray} 
\Delta \Gammati &=& \delta d \Gamma - \delta \big( d(J^{-1}) \wedge dJ \big) + d(J^{-1} A ), \label{eqn1} \\
\Delta J &=& \delta ( J \mm \Gamma ) - \langle d J ; \tilde{\Gamma}\rangle - A , \label{eqn2} \\
d \vec{A} &=& \overrightarrow{\text{div}} \big(dJ \wedge \Gamma\big) + \overrightarrow{\text{div}} \big( J\, d\Gamma\big) - d\big(\overrightarrow{\langle d J ; \tilde{\Gamma}\rangle }\big),   \label{eqn3}\\
\delta \vec{A} &=& v,  \label{eqn4}
\end{eqnarray}
with boundary data 
\begin{eqnarray}   
Curl(J) \equiv \partial_j J^\mu_i - \partial_i J^\mu_j =0 \ \ \text{on} \ \partial \Omega,  \label{bdd1}
\end{eqnarray}
where $v\in W^{m-1,p}(\Omega)$ is some vector valued $0$-form free to be chosen.
Then for each $p\in\Omega$,  there exists a neighborhood $\Omega'\subset\Omega$ of $p$ such that $J$ is the Jacobian of a coordinate transformation $x \mapsto y$ on $\Omega'$, and the components of $\Gamma$ in $y$-coordinates are in $W^{m+1,p}(\Omega')$. \vspace{.15cm} \newline 
Conversely, if there exists a coordinate transformation $x \mapsto y$ with Jacobian $J = \frac{\partial y}{\partial x} \in W^{m+1,p}(\Omega)$ such that the components of $\Gamma$ in $y$-coordinates are in $W^{m+1,p}(\Omega)$, then there exists $\Gammati \in W^{m+1,p}(\Omega)$ and $A\in W^{m,p}(\Omega)$ such that $(J,\Gammati,A)$ solve \eqref{eqn1} - \eqref{bdd1} in $\Omega$ for some $v\in W^{m-1,p}(\Omega)$.
\end{Thm}

Equations (\ref{eqn1})-(\ref{eqn4}) are the RT-equations. To clarify the choice of Sobolev space, note that $L^p$ is not closed under products, and as a result the regularity $\Gamma,\Riem(\Gamma)\in W^{m,p}$, $m\geq1$, $p>n,$ is assumed in the proof of convergence of the iteration scheme in \cite{ReintjesTemple_ell2} to control the nonlinear products on the right hand side of the RT-equations (\ref{eqn1})-(\ref{eqn4}), $m\geq1$, $p>n$ being the lowest regularity which implies $\Gamma, Riem(\Gamma)$ are H\"older continuous by Morrey's inequality.\footnote{Controlling the nonlinear products is a main obstacle to extending the existence theory for the RT-equations to connections of lower regularity, say $L^p,$ or $m<1.$ Alternatively, assuming connections in $L^{\infty}\subset L^p$, a natural setting of GR shock waves, is sufficient to control the nonlinear products, but is problematic due to the existence of Calder\'{o}n-Zygmund singularities.}    

In Section \ref{Sec_Prelim} we introduce the {\it vectorization} of $A,$ denoted $\vec{A}$, as the vector valued $1$-form $\vec{A} \equiv A^{\mu}_{i}dx^i$, so $d\vec{A} = Curl (A)$. Equation \eqref{eqn3} is then obtained by setting $d$ of the vectorized right hand side of (\ref{eqn2}) equal to zero, so the identity $d\vec{A} = Curl (A)$ implies that \eqref{eqn3}  is equivalent to the integrability condition $Curl( J)=0$ for the Jacobian, c.f. Lemma \ref{Lemma_integrability} below. The operations $\vec{\cdot},$ $\overrightarrow{\rm div}$ and $\langle \cdot \; \cdot \rangle$, introduced in Sections \ref{Sec_Prelim}, are special operations on matrix valued differential forms meaningful when the dimension of the matrices equals the dimension of the physical space. New features arise in the auxiliary Euclidean Cartan algebra essentially because we view $J$ both as a matrix valued zero form and a vector valued $1$-form at different stages of the argument.  This framework, which bridges matrix and vector valued differential forms through special operations, appears to be forced on us to close the equations.

The vector valued $0$-form $v$ in the RT-equations (\ref{eqn1})-(\ref{eqn4}) is free to be chosen, and reflects the freedom in the problem to apply sufficiently smooth coordinate transformations, which preserve optimal metric regularity. Equation \eqref{eqn4} has been introduced so that (\ref{eqn3}) - (\ref{eqn4}) take the Cauchy-Riemann form $d\vec{A}=f$, $\delta\vec{A}=g$.   Such systems require the consistency conditions $df=0$, $\delta g=0$, (c.f. Section \ref{Sec_Cauchy-Riem}).   Condition $df=0$ is met in (\ref{eqn3}) because the derivation shows the right hand side is exact, (again, equation (\ref{eqn3}) is obtained by setting $d$ of the vectorized right hand side of (\ref{eqn2}) equal to zero), and $\delta g=0$ in (\ref{eqn4}) because $\delta v=0$ holds as an identity for $0$-forms.     

A crucial point in the derivation of the RT-equations is the regularization of the term $d \big(\overrightarrow{\delta ( J \mm \Gamma )}\big) $. This term appears when we take $d$ of the right hand side of \eqref{eqn2} to derive \eqref{eqn3}, and appears to be one derivative too low to get the required regularity $A\in W^{m,p}$.  That is,  our assumptions control $d\Gamma$ in $W^{m,p}$, but not $\delta\Gamma$ which measures the derivatives not controlled by $d\Gamma$, c.f. \eqref{Gaffney}.  The regularity $A\in W^{m,p}$ is required for \eqref{eqn1}  and  \eqref{eqn2} to imply the sought after regularity for $J$ and $\Gammati$.  Surprisingly,  $d \big(\overrightarrow{\delta ( J \mm \Gamma )}\big)$ is one order smoother than it initially appears to be because it
 can (and has been) re-expressed in terms of $d\Gamma$ via the identity $ d\big(\overrightarrow{\delta ( J \mm \Gamma )}\big)  = \overrightarrow{\text{div}} \big(dJ \wedge \Gamma\big) + \overrightarrow{\text{div}} \big( J d\Gamma\big) $, (see Lemma \ref{Lemma_regularity-miracle} below).  This extra derivative ``miracle'', necessary for the consistency of \eqref{eqn1} - \eqref{eqn4} with the Riemann-flat condition, made us believe that resolving the problem of optimal regularity in GR by the RT-equations would work.

The second order system (\ref{eqn1}) - \eqref{eqn2} comes from first order Cauchy-Riemann equations equivalent to the Riemann-flat condition $Riem(\Gamma-\Gammati)=0$. The advantage of the second order system over the first order system is that it gives us the freedom to solve (\ref{eqn1}) with arbitrary boundary conditions, and \eqref{eqn2} with the boundary data \eqref{bdd1}, without being forced to use problematic implicit boundary data of the form \eqref{Cauchy_Riem_bd_smooth} which is required for the standard equivalence between solutions of Cauchy-Riemann and Poisson type equations, c.f. Section \ref{Sec_Cauchy-Riem}.   This freedom of assigning boundary conditions is a propitious feature of the RT-equations.         

We end the section with an overview of the derivation of the RT-equations below. The idea is to view the Riemann-flat condition ${\it Riem}(\Gamma-\Gammati)=0$  as an equation for $d\Gammati,$ namely, $d\Gammati=d\Gamma+(\Gamma-\Gammati)\wedge(\Gamma-\Gammati),$ and then to augment this to a first order system of Cauchy-Riemann equations by addition of the equation for $\delta\Gammati=h$ with arbitrary $h$.  But to obtain a solvable system, it is necessary to couple this Cauchy-Riemann system in the unknown $\Gammati$ to an equation in the unknown Jacobian $J$. For this, we use an equivalent form of the Riemann-flat condition, $J^{-1}dJ=\Gamma-\Gammati$. Now $d\Gammati=d\Gamma+(\Gamma-\Gammati) \wedge(\Gamma-\Gammati)$ and $J^{-1}dJ=\Gamma-\Gammati$ are not independent, both being equivalent to the Riemann-flat condition.   To obtain two independent equations, we next employ the identity $d\delta+\delta d=\Delta$ to derive two semi-linear elliptic Poisson equations, one for $\Delta\Gammati$, and one for $\Delta J$. This results in the two second order equations \eqref{eqn1} - \eqref{eqn2}, which closes in $(\Gammati,J)$ for fixed $A$. The equations are formally correct at the levels of regularity sufficient for $\Gammati$ to be one order smoother than $\Gamma$, consistent with known results on elliptic smoothing by the Poisson equation in $L^p$-spaces, \cite{Dac,Evans,Grisvard}. To impose the integrability condition for $J$, we use the freedom in $\delta\Gammati$ to introduce the variable $A$ into the right hand sides of \eqref{eqn1}  and \eqref{eqn2}, and impose $Curl\vec{J}=0$ by asking that the equation in $A$ be obtained by requiring $d$ of the vectorized right hand side of the $J$ equation \eqref{eqn2}, be zero.    This is \eqref{eqn3}. To obtain the final form of the RT-equations, we apply our fortuitous identity by which all {\it bad terms} in the equations involving $\delta\Gamma$ can be re-expressed in terms of $d\Gamma$, leading to a gain of one derivative on the right hand side.    Equation \eqref{eqn4} then represents the ``gauge freedom'' to impose $\delta A=v$, and the boundary condition \eqref{bdd1} simply imposes the integrability condition for $J$ on the boundary.    The resulting system \eqref{eqn1}-\eqref{bdd1}, the RT-equations, requires no boundary condition for $\Gammati$ or for $A$, only the boundary condition \eqref{bdd1} for $J$.   
Now the regularity on the right hand sides of the RT-equations imply that solutions $\Gammati,J$ of the RT-equations provide $\Gammati$ with one more derivative than $\Gamma$, but it turns out that $\Gammati$ need not satisfy the Riemann-flat condition because the RT-equations have a larger solution space than the first order equations from which they are derived. To complete the forward implication in Theorem \ref{Thm_main}, we define $\Gammati' \equiv \Gamma-J^{-1}dJ$ which meets the Riemann-flat condition by definition but an additional argument based on elliptic regularity is required to show that $\Gammati'$ is indeed one level smoother than $\Gamma$.                

In summary, we start with two equivalent first order equations, one for $d\Gammati$ and one for $dJ$,  both equivalent to the Riemann-flat condition.   Out of these, we create two independent nonlinear Poisson equations in $\Gammati$ and $J$, and the system has the freedom to impose an auxiliary solution $A$ through the gauge freedom to impose $\delta\Gammati$.  Not all solutions of the RT-equations provide a $\Gammati$ which solves the Riemann-flat condition, but given any solution $(\Gammati,J,A)$ of \eqref{eqn1} - \eqref{bdd1}, we show that there is enough freedom in $A$ such that there always exists $A'$ such that $(\Gammati',J,A'),$ with $\Gammati'=\Gamma-J^{-1}J$, also solves the RT-equations, and $\Gammati'$ meets the Riemann-flat condition by construction.  Now the RT-equations are formally consistent with smoothing according to the linear theory of elliptic smoothing in $L^p$ spaces, but the RT-equations are nonlinear, and the boundary data (\ref{bdd1}) is non-standard.  Thus an existence theory based on finding a suitable convergent iteration with modified initial data, is required to reduce the existence theory to known theorems on linear elliptic PDE's, and to thereby establish that the whole theory actually works.  This is accomplished for the first time in  \cite{ReintjesTemple_ell2}.
 
In Section \ref{Sec_Prelim} we introduce the auxiliary Euclidean Cartan algebra of matrix-valued $k$-forms, and define the operations $\overrightarrow{\rm div}$ and $\langle\cdot\: ; \cdot \rangle$, (c.f. (\ref{eqn3}) and (\ref{eqn2}), respectively). In Section \ref{Sec_Riem-flat}, we express the Riemann-flat condition within the framework of matrix and vector valued differential forms.  In Section \ref{Sec_Cauchy-Riem} we clarify the connection between the first order Cauchy-Riemann equations and the Poisson equation in the setting of matrix valued differential forms. In Sections \ref{Sec_equiv_warm-up} and \eqref{Sec_equiv_main} we derive the RT-equations (\ref{eqn1})-(\ref{eqn4}) together with an alternative formulation (in Section \ref{Sec_equiv_system}), and prove our main result, Theorem \ref{Thm_main}.

\section{Matrix valued differential forms}  \label{Sec_Prelim}

In this section we develop a theory of matrix valued differential forms in the special case when the dimension of the matrix components agrees with the dimension of the space, $n$.   The exterior derivative $d$ and its co-derivative $\delta$ operate on matrix valued $k$-forms component-wise,  and the wedge product introduces the matrix commutator, both of which are independent of the size of the matrices.  However, to close the equations, we need to introduce two new operations, c.f. \eqref{eqn4}.   The first operation maps matrix valued $0$-forms $A$ to vector valued $1$-forms $\vec{A}$ via contraction of one matrix indices with $dx^i$.  The second is a vectorized divergence $\vec{div}$ which maps matrix valued  $k$-forms to vector valued $k$-forms by taking the divergence with respect to the lower matrix index.   These vectorizing operations are meaningful only for matrix valued forms in which the matrices and the dimension of the space are both equal.

Keep in mind, this is a {\it Euclidean} framework because we only consider matrix valued differential forms in the fixed coordinate system $x$ in which our connection $\Gamma\equiv \Gamma^k_{ij}$ is originally assumed to be given, and we take the auxiliary metric on $x$ to be Euclidean.  Since $x$ is assumed fixed,  the covariance properties of these differential forms is not an issue.  

To start, we interpret the connection $\Gamma$ as a matrix valued $1$-form $\Gamma^\mu_\nu \equiv \Gamma^\mu_{\nu i} dx^i$, in which case the Riemann curvature tensor of $\Gamma$ can be written as the matrix valued $2$-form
\beq \label{Riemann_2-form} 
 {\rm Riem}(\Gamma) = d \Gamma + \Gamma \wedge \Gamma,
\eeq
c.f., Lemma \ref{Lemma_Riemann_2-form}.  By a matrix valued differential $k$-form $A$ we mean an $(n\times n)$-matrix whose components are $k$-forms over $n$-dimensional base space $\Omega\subset \R^n$, and we write
\beq \nonumber
A = A_{[i_1...i_k]} dx^{i_1} \wedge ... \wedge dx^{i_k} \equiv \sum_{i_1< ... < i_k} A_{i_1...i_k} dx^{i_1} \wedge ... \wedge dx^{i_k},
\eeq 
for $(n\times n)$-matrices  $A_{i_1...i_k}$  that are totally anti-symmetric in the indices $i_1,...,i_k \in \{1,...,n\}$.   As is standard, we always indicate an increasing ordering of indices by a square bracket around the indices and we set 
\beq \label{def_wedge_basis}
dx^{i_1} \wedge ... \wedge dx^{i_k} \equiv  \sum_{\pi \in S_k} {\rm sgn}(\pi)\: dx^{i_{\pi(1)}} \otimes ... \otimes dx^{i_{\pi(k)}},
\eeq
where $S_k$ denotes the set of all permutations of $\{1,...,k\}$. We define the exterior derivative of a matrix valued $k$-form by 
\begin{eqnarray} \label{def_exterior_deriv}
d A &\equiv&   d\big( A_{[i_1...i_k]} \big) \wedge dx^{i_1} \wedge ... \wedge dx^{i_k}  \cr
&=&  \partial_l A_{[i_1...i_k]} dx^l \wedge dx^{i_1} \wedge ... \wedge dx^{i_k} ,
\end{eqnarray}
and we define the wedge product of a matrix valued $k$-form $A$ with a matrix valued $l$-form $B = B_{j_1...j_l} dx^{j_1} \wedge ... \wedge dx^{j_l}$ as  
\begin{eqnarray} \label{def_wedge}
A \wedge B  
&\equiv & \frac{1}{l!k!} A_{i_1...i_k} \mm B_{j_1...j_l} \; dx^{i_1} \wedge ... \wedge dx^{i_k} \wedge dx^{j_1} \wedge ... \wedge dx^{j_l}, 
\end{eqnarray}
where the dot denotes standard matrix multiplication. The wedge product of a matrix valued $k$-form with itself is in general non-zero unless the component matrices commute. In fact, for any matrix valued $1$-forms $A=A_i dx^i$ commutativity of its component matrices is a necessary and sufficient condition  for $A\wedge A=0$, as we now show by computing
\begin{eqnarray} \label{wedge_non-zero}
(A_i dx^i)\wedge (A_j dx^j)  
 &\equiv & A_i \mm A_j dx^i\wedge dx^j \cr 
 &=&  A_i \mm A_j \: \big( dx^i\otimes dx^j - dx^j\otimes dx^i \big) \cr
 &=&  (A_i \mm A_j - A_j \mm A_i) \: dx^i\otimes dx^j 
\end{eqnarray}
and this vanishes if and only if  $A_i A_j - A_j A_i=0$. The main point is that $\Gamma \wedge \Gamma$ in \eqref{Riemann_2-form} is in general non-vanishing. 

To define the co-derivative $\delta$ and the Laplace operator $\Delta$ for matrix valued $k$-forms, define the Hodge star operator $*$ by
\beq \label{def_Hodge}
A \wedge (*B)   \equiv \langle A\: ;B\rangle  dx^1 \wedge ... \wedge dx^n ,
\eeq
for matrix valued $k$-forms $A$ and $B$, where we define
\beq \label{def_inner-product}
\langle A\; ; B \rangle^\mu_\nu \equiv \sum_{i_1<...<i_k} A^\mu_{\sigma\: i_1...i_k} B^\sigma_{\nu\: i_1...i_k} .
\eeq 
The matrix valued inner product \eqref{def_inner-product} generalizes the Euclidean inner product on the components of $k$-forms; \eqref{def_inner-product} is symmetric in $A$ and $B$ if and only if $A$ and $B$ commute and for a matrix valued $0$-form $J$ we have
\beq \label{inner-product_muliplications}
 J \cdot \langle A\; ; B \rangle =  \langle  J \mm A\; ; B \rangle 
 \hspace{.5cm} \text{and}  \hspace{.5cm}  
\langle A\mm J \; ; B \rangle =  \langle A\; ;  J \mm B \rangle .
\eeq   
Now, the Hodge-star operator $*$ maps $k$-forms linearly to $(n-k)$-forms and commutes with matrices, $*(J\mm B) = J\mm  (*B)$ for any matrix $J$, since 
\begin{eqnarray}\nonumber
A \wedge *(J B)  &=& \langle A\: ; J B\rangle \: dx^1 \wedge ... \wedge dx^n \cr
&=& \langle A J\: ; B\rangle \: dx^1 \wedge ... \wedge dx^n \cr
&=& A J \wedge * B 
\end{eqnarray}
by \eqref{def_Hodge} and \eqref{inner-product_muliplications}. Moreover, \eqref{def_Hodge} is equivalent to the orthogonality condition (for increasing indices)
\beq \label{def_Hodge_equiv}
dx^{[i_1} \wedge ... \wedge dx^{i_k]} \wedge * \big( dx^{[j_1} \wedge ... \wedge dx^{j_k]} \big) 
= \begin{cases} dx^1 \wedge ... \wedge dx^n, \hspace{.3cm} \text{if} \ \ i_1=j_1,...,i_k=j_k, \cr 0 \hspace{.3cm} \text{otherwise},    \end{cases}
\eeq 
since we find from definition \eqref{def_wedge} of the wedge product that 
\begin{eqnarray} \nonumber
\big( A \wedge (*B) \big)^\mu_\nu  
&=& A^\mu_{\sigma [i_1... i_k]} \: B^\sigma_{\nu [j_1... j_k]} \: dx^{i_1} \wedge ... \wedge dx^{i_k} \wedge * \big( dx^{j_1} \wedge ... \wedge dx^{j_k} \big)  ,
\end{eqnarray}
while we find from definition \eqref{def_Hodge} of the Hodge-star that
\begin{eqnarray} \nonumber
\big( A \wedge (*B) \big)^\mu_\nu  &=& \langle A\: ;B\rangle^\mu_\nu \; dx^1 \wedge ... \wedge dx^n.
\end{eqnarray}
so that \eqref{def_Hodge_equiv} follows directly by comparing coefficients. Now, by \eqref{def_Hodge_equiv}, the Hodge star maps a basis element to its complementary element, from which we find that $**A= (-1)^{k(n-k)}A$, where the factor $(-1)^{k(n-k)}$ appears when passing the dual basis element to the left hand side, and so 
$$*^{-1} = (-1)^{k(n-k)} *.$$ 
The co-derivative of a $k$-form $A$ is now defined (in the standard way) as the $(k-1)$-form 
\beq \label{def_coderiv}
\delta A \equiv (-1)^{n-k} \: * \big( d (*^{-1} A) \big)
\eeq
and the Laplace operator as 
\beq \label{def_Laplacian}
\Delta  \equiv \delta d + d \delta.
\eeq
The Laplacian acts on each component of A as the scalar Laplacian, 
\beq \label{basics_Laplacian}
(\Delta A)^\mu_{\nu i_1...i_k} 
= \Delta \big(A^\mu_{\nu i_1...i_k}\big) 
= \sum_{j=1}^n \partial_{j}\partial_{j}\big(A^\mu_{\nu i_1...i_k}\big),
\eeq
c.f. Theorem 3.7 in \cite{Dac}, (where the last identity in \eqref{basics_Laplacian} holds when $x^i$ are Euclidean coordinates, the case we have here). A straightforward computation shows that $\delta A=0$ for $0$-forms $A$, and if $k=1$, then the co-derivative is the divergence,
\beq \label{basics_delta}
(\delta A)^\mu_\nu = \sum_{i=1}^n \partial_i A^\mu_{\nu\: i} .
\eeq             

With the exception of property \eqref{wedge_non-zero} of the wedge product, matrix valued differential forms behave like standard scalar differential forms with scalar multiplication replaced by matrix multiplication whenever components are multiplied.  In particular, the derivative operations \eqref{def_exterior_deriv}, \eqref{def_coderiv} and \eqref{def_Laplacian} simply act component-wise on matrix components. We now prove that \eqref{Riemann_2-form} holds for the Riemann curvature tensor
\beq \nonumber
\Riem(\Gamma)^\mu_\nu \equiv R^{\mu}_{\nu ij}dx^i \otimes dx^j,
\eeq
the components of which are given by
\beq \label{Riemann_comp}
\Riem(\Gamma)^{\mu}_{\nu ij} \equiv R^{\mu}_{\nu ij} \equiv \Gamma^{\mu}_{\nu j,i}-\Gamma^{\mu}_{\nu i,j}+\: \Gamma^{\mu}_{\sigma i}\Gamma^{\sigma}_{\nu j}-\Gamma^{\mu}_{\sigma j}\Gamma^{\sigma}_{\nu i} 
\eeq
and where we interpret $\mu$ and $\nu$ as matrix indices.

\begin{Lemma}   \label{Lemma_Riemann_2-form}
In fixed coordinates $x^i$, the Riemann curvature tensor is the matrix-valued $2$-form \eqref{Riemann_2-form} with matrix components 
\beq \label{Riemann_2-form_comp}
\Riem(\Gamma)^\mu_\nu 
\: =\: R^{\mu}_{\nu [ij]}dx^i \wedge dx^j
\: =\:  d\big(\Gamma^{\mu}_{\nu i}dx^i\big)  +\:  \Gamma^{\mu}_{\sigma i} dx^i \:  \wedge \: \Gamma^{\sigma}_{\nu j} dx^j .
\eeq
\end{Lemma}       

\Proof
We use \eqref{def_wedge_basis} and the antisymmetry of $R^\mu_{\nu ij}$ in $i$ and $j$ to write  
\begin{eqnarray} \nonumber
 R^{\mu}_{\nu [ij]}\: dx^i\wedge dx^j  
 &=&   R^{\mu}_{\nu [ij]} \:\big( dx^i\otimes dx^j - dx^j\otimes dx^i \big) \cr
  &=& \sum_{i<j} R^{\mu}_{\nu ij}  dx^i\otimes dx^j +    \sum_{i<j} R^{\mu}_{\nu ji} dx^j\otimes dx^i \cr
&=&  R^{\mu}_{\nu ij}dx^i\otimes dx^j ,
\end{eqnarray}
without losing any information of the curvature tensor, which turns ${\rm Riem}(\Gamma)$ into a matrix valued $2$-form.  
To prove the second equality in \eqref{Riemann_2-form_comp}, use \eqref{def_exterior_deriv} to compute
\begin{eqnarray} \nonumber
d\big(\Gamma^{\mu}_{\nu i}dx^i\big) 
&=& \Gamma^{\mu}_{\nu i, j} dx^j \wedge dx^i   
\: = \: \Gamma^{\mu}_{\nu i, j} \big(dx^j \otimes dx^i - dx^i \otimes dx^j \big) \cr
&=&  \big(\Gamma^{\mu}_{\nu j,i}-\Gamma^{\mu}_{\nu i,j}\big) dx^i\otimes dx^j 
\end{eqnarray}
and use \eqref{def_wedge} to compute
\begin{eqnarray} \nonumber
\Gamma^{\mu}_{\sigma i} dx^i \:  \wedge \: \Gamma^{\sigma}_{\nu j} dx^j
& =&  \Gamma^{\mu}_{\sigma i} \Gamma^{\sigma}_{\nu j} \: dx^i \wedge dx^j \: 
=\: \Gamma^{\mu}_{\sigma i} \Gamma^{\sigma}_{\nu j} \big(dx^i\otimes dx^j -dx^j\otimes dx^i\big)  \cr
&=&  \big( \Gamma^{\mu}_{\sigma i} \Gamma^{\sigma}_{\nu j} - \Gamma^{\mu}_{\sigma j} \Gamma^{\sigma}_{\nu i}\big) dx^i \otimes dx^j 
\end{eqnarray}
which combined yields the sought after second equality in \eqref{Riemann_2-form_comp}.
\QED

To proceed, let $W^{m,p}(\Omega)$ be the Sobolev space of functions with partial derivatives up to $m$-th order in $L^p$. We say that a matrix valued $k$-form $w$ is in $W^{m,p}(\Omega)$ if its components are functions in $W^{m,p}(\Omega)$, with respect to the fixed coordinate system $x$. Assume now that $m\geq 1$ and $p>n$, so that the Sobolev embedding theorem implies functions in $W^{1,p}$ are H\"older continuous, c.f. Morrey's inequality in \cite{Evans}. The following Leibnitz rule holds.        

\begin{Lemma} \label{Lemma_Leibnitz-rule}
Let $A\in W^{1,p}(\Omega)$ be a matrix valued $k$-form and let $B \in W^{1,p}(\Omega)$ be a matrix valued $j$-form, and assume $p>n$, then 
\beq \label{Leibniz_rule} 
d(A\wedge B) = dA \wedge B + (-1)^k A \wedge dB \ \ \ \in \ L^p(\Omega).
\eeq
\end{Lemma}                      

\Proof
Assuming first that $A$ and $B$ are smooth, a straightforward computation gives
\begin{eqnarray}  \label{techeqn0_Leibnitz-rule}
d(A\mm B)^\mu_\nu 
&=& \tfrac{1}{l!k!} d\big( A^\mu_{\sigma i_1 ... i_k}  B^\sigma_{\nu j_1 ... j_l}\: dx^{i_1}\wedge ... \wedge dx^{i_k} \wedge dx^{j_1}\wedge ... \wedge dx^{j_l} \big) \cr
&=& \tfrac{1}{l!k!} \partial_l \big( A^\mu_{\sigma i_1...i_k} B^\sigma_{\nu j_1 ... j_l}\big) dx^l\wedge dx^{i_1}\wedge...\wedge dx^{i_k} \wedge dx^{j_1}\wedge...\wedge dx^{j_l}  \cr
&=&  \partial_l A^\mu_{\sigma [i_1...i_k]} dx^l\wedge dx^{i_1}\wedge ... \wedge dx^{i_k} \wedge  B^\sigma_{\nu} \cr
&& +\:  A^\mu_{\sigma} \, \wedge \, (-1)^k \partial_l B^\sigma_{\nu[j_1 ... j_l]}  dx^l\wedge dx^{j_1}\wedge ... \wedge dx^{j_l} \cr
&=& dA^\mu_\sigma \wedge B^\sigma_\nu + (-1)^k A^\mu_\sigma\, dB^\sigma_\nu,
\end{eqnarray}
which is the sought after identity \eqref{Leibniz_rule}. To extend the above computation to $W^{1,p}$, we use that the difference quotient (along the $j$-th coordinate axis) $D_{h} f$ of a function $f \in W^{1,p}(\Omega)$ converges to its weak derivative $\partial_j f$ in $L^1$ as $h\rightarrow 0$. It follows that for the product of two functions $f,g\in W^{1,p}(\Omega)$ we have at $x\in \Omega$
\beq \label{techeqn1_Leibnitz-rule}
D_h(fg)|_x= D_h(f)|_x \: g(x+h) + f(x) D_h(g)|_x.
\eeq
Now, since $p>n$, we know by the Sobolev embedding theorem that $g$ and $f$ are H\"older continuous, so that the right hand side in \eqref{techeqn1_Leibnitz-rule} converges in $L^1$ as $h\rightarrow 0$ and implies
\beq \nonumber
\lim_{h\rightarrow 0} D_h(fg) = g\: \partial_j f  + f\: \partial_j g \in L^p(\Omega).
\eeq
Thus, since $D_h(fg)$ converges to the weak derivative $\partial_j(fg)$ in $L^1$ as $h\rightarrow 0$, we  conclude that
\beq \label{techeqn2_Leibnitz-rule}
\partial_j (fg) = g\: \partial_j f + f\: \partial_j g \in L^p(\Omega)
\eeq
and thus $fg\in W^{1,p}(\Omega)$. Applying \eqref{techeqn2_Leibnitz-rule} component-wise for the third equality in \eqref{techeqn0_Leibnitz-rule} leads to the sought after equation \eqref{Leibniz_rule}. 
\QED

We also require the following Leibnitz rule for the co-derivative.

\begin{Lemma} \label{Lemma_colon}
Let $J\in W^{2,p}(\Omega)$ be a matrix valued $0$-form and~$w \in W^{2,p}(\Omega)$ a matrix valued $1$-form, then
\beq \label{colon}
\delta (J\mm w ) = J \mm \delta w  + \langle d J ;  w \rangle
\eeq
where $\langle\cdot ; \cdot \rangle$ is the matrix valued inner product defined in \eqref{def_inner-product}.
\end{Lemma}

\Proof
Using that $\delta$ of a $1$-form is the divergence \eqref{basics_delta}, we find that
\beq \nonumber
\big(\delta ( J \mm w ) \big)^\alpha_i = \delta \big( J^\alpha_k w^k_{ij} dx^j \big) = \sum_{j=1}^n \partial_j \big( J^\alpha_k w^k_{ij}  \big)  = \sum_{j=1}^n  J^\alpha_{k,j} w^k_{ij} +  J^\alpha_k (\delta w)^k_{i} 
\eeq
and this proves the lemma.
\QED

We close this section by introducing the two operations we require to close the equations, which relate matrix valued to vector valued differential forms. Note, a matrix valued $0$-form $J^\alpha_i$ turns into a vector valued $1$-form $J^\alpha_i dx^i$ by contracting the lower matrix index with a Cartan basis element, (where $\alpha$ labels the components of the vector). To start, let an arrow over a matrix valued $0$-form $A$ convert  $A$ to its equivalent vector valued $1$-form, i.e., 
\beq\label{defnarrow1}
\vec{A}\equiv A^\alpha_i dx^i.
\eeq 
By this, we can express the integrability of the Jacobian $J$, (c.f., Frobenius Theorem, \cite{Taylor}),
as
\beq \label{J_integrability_intro}
d\vec{J} =0,
\eeq
since                
\beq \nonumber 
Curl(J)\equiv \frac12 \big( J^\alpha_{i,j} - J^\alpha_{j,i} \big) dx^j \otimes dx^i = J^\alpha_{i,j} dx^j\wedge dx^i =  d(J^\alpha_i dx^i) \equiv  d\vec{J}^\alpha.
\eeq 
  For our elliptic system to close, we need one more operation to convert matrix valued to vector valued differential forms. Namely, for $\omega \in \Lambda^{1,p}_k(\Omega)$, we define 
\beq \label{Def_vec-div}
\overrightarrow{\text{div}}(\omega)^\alpha \equiv \sum_{l=1}^n \partial_l \big( (\omega^\alpha_l)_{i_1,,,i_k}\big) dx^{i_1}\wedge . . . \wedge dx^{i_k} ,
\eeq
which is the divergence with respect to the lower matrix index, thus creating a vector valued $k$-form out of a matrix valued $k$-form. We close this subsection with the following intriguing identity for commuting $d$ and $\delta$ which has no analogue for classical scalar valued differential forms and is the key identity for the regularity of the final elliptic system to close.

\begin{Lemma} \label{Lemma_regularity-miracle}
Let $\Gamma \in W^{m,p}(\Omega)$ and $J \in W^{m+1,p}(\Omega)$ for $p>n$ and $m\geq1$, then
\beq \label{regularity-miracle}
d \big(\overrightarrow{\delta ( J \mm \Gamma )}\big) 
= \overrightarrow{\text{div}} \big(dJ \wedge \Gamma\big) + \overrightarrow{\text{div}} \big( J\mm d\Gamma\big) .
\eeq
\end{Lemma}

\Proof
Since $\delta$ of a matrix valued $1$-form is the divergence (for fixed matrix components), we have 
\beq \nonumber
\big(\delta ( J \Gamma )\big)^\alpha_j 
= \delta \big(J^\alpha_k \Gamma^k_{ji} dx^i  \big) 
= \sum_{l=1}^n \partial_l \big( J^\alpha_k \Gamma^k_{jl} \big) 
\eeq 
and thus
\beq \nonumber
\big(\overrightarrow{\delta ( J \Gamma )}\big)^\alpha 
= \big(\delta ( J \Gamma )\big)^\alpha_j dx^j 
=  \sum_{l=1}^n \partial_l \big( J^\alpha_k \Gamma^k_{jl} \big) dx^j,
\eeq 
from which we find that
\beq \nonumber
d\big(\overrightarrow{\delta ( J \Gamma )}\big)^\alpha 
= \partial_i \big(\delta ( J \Gamma )\big)^\alpha_j dx^i\wedge dx^j
= \sum_{l=1}^n \partial_i\partial_l \big( J^\alpha_k \Gamma^k_{jl} \big) dx^i\wedge dx^j,
\eeq 
where in the case $m=1$ second derivatives are taken in a distributional sense. Now, since weak derivatives commute, we obtain from the product rule (which applies since $\Gamma$ and derivatives of $J$ are H\"older continuous) that
\begin{eqnarray} \nonumber
d\big(\overrightarrow{\delta ( J \Gamma )}\big)^\alpha 
&=& \sum_{l=1}^n \partial_l \partial_i \big( J^\alpha_k \Gamma^k_{jl} \big) dx^i\wedge dx^j \cr
&=& \sum_{l=1}^n \partial_l \big( J^\alpha_{k,i}  \Gamma^k_{jl} \big)  dx^i\wedge dx^j 
+ \sum_{l=1}^n \partial_l \big( J^\alpha_k \Gamma^k_{lj,i} \big) dx^i\wedge dx^j \cr
&=& \overrightarrow{\text{div}} \big(dJ \wedge \Gamma\big)^\alpha + \overrightarrow{\text{div}} \big( J\mm d\Gamma\big)^\alpha .
\end{eqnarray} 
This completes the proof. 
\QED

\section{The Riemann-flat condition in terms of matrix valued differential forms} \label{Sec_Riem-flat}

Consider the transformation law for a connection  
\beq \label{prelim_connection_transfo_1}
(J^{-1})^k_\alpha \big( \partial_j J^\alpha_{i} + J^\beta_i J^\gamma_j \Gamma^\alpha_{\beta\gamma} \big) = \Gamma^k_{ij} ,
\eeq
where $\Gamma^k_{ij}$ denotes the components of the connection in $x^i$-coordinates,  $\Gamma^\alpha_{\gamma\beta}$ denotes its components in $y^{\alpha}$-coordinates, and where $J^\alpha_i=\frac{\partial y^\alpha}{\partial x^i}$ denotes the  Jacobian of the coordinate transformation. 
Assume now that $\Gamma^k_{ij} \in W^{m,p}(\Omega)$, $J^\alpha_i \in W^{m+1,p}(\Omega)$ and $\Gamma^\alpha_{\gamma\beta} \in W^{m+1,p}(\Omega)$, for $m\geq 1$, so the Jacobian $J$ smooths the connection $\Gamma^k_{ij}$ by one order.  For these given coordinates $x$ and $y$, define
\beq \label{def_tensor_Gammati}
\tilde{\Gamma}^k_{ij} \equiv (J^{-1})^k_\alpha  J^\beta_i J^\gamma_j \Gamma^\alpha_{\beta\gamma}.
\eeq
Then requiring $\Gammati$ to transform as a $(1,2)$-tensor, \eqref{def_tensor_Gammati} defines the {\it tensor} $\Gammati$.  By this, \eqref{prelim_connection_transfo_1} can be written equivalently as 
\beq \label{prelim_connection_transfo_2}
(J^{-1})^k_\alpha \; \partial_j J^\alpha_{i} = (\Gamma-\tilde{\Gamma})^k_{ij}.
\eeq
Now, since adding a tensor to a connection yields another connection, \eqref{prelim_connection_transfo_2} is just the condition that $J$ transforms the connection $\Gamma-\tilde{\Gamma}$ to zero.  This implies $\Gamma-\tilde{\Gamma}$ is a Riemann-flat connection, $
{\rm Riem} (\Gamma-\tilde{\Gamma}) =0.$ In the language of matrix valued differential forms (\ref{prelim_connection_transfo_2}) reads
\begin{eqnarray} \label{SmoothingCondition}
J^{-1}dJ &=&\Gamma-\tilde{\Gamma},
\end{eqnarray}
where $J$ is a matrix valued $0$-form and $\Gamma$ and $\tilde{\Gamma}$ are matrix valued $1$-forms.  Equation \eqref{SmoothingCondition} plays a central role in this paper. 

Note \eqref{prelim_connection_transfo_1} - \eqref{prelim_connection_transfo_2} apply to $\Gamma^k_{ij} \in L^\infty(\Omega)$ and $\Gamma^\alpha_{\gamma\beta} \in C^{0,1}(\Omega)$, and it is proven in \cite{ReintjesTemple_geo} that the reverse implication is also true, even at this low level of regularity of $\Gamma\in L^{\infty}$.   The equivalence is this:
One can smooth an $L^\infty$ connection $\Gamma$ one order to $C^{0,1}$ by a $C^{0,1}$ coordinate transformation if and only if the \emph{Riemann-flat condition} holds, and we say that the Riemann-flat condition holds if there exists a Lipschitz continuous rank $(1,2)$-tensor $\Gammati^k_{ij}$ with symmetry $\Gammati^k_{ij}=\Gammati^k_{ji}$ such that $Riem(\Gamma-\Gammati)=0$ holds. (In light of Theorem \eqref{Thm_Prelim} we sometimes also refer to \eqref{SmoothingCondition} as the Riemann-flat condition.) Based on this, we now record the following version of the Riemann flat condition and related equivalencies applicable to the smoothness classes $\Gamma\in W^{m,p}$ relevant for this paper.

\begin{Thm} \label{Thm_Prelim}
Let $\Gamma^k_{ij}$ be a symmetric connection in $W^{m,p}(\Omega)$ for $m\geq 1$ and $p>n$  (in coordinates $x^i$).\footnote{Note that symmetry of $\Gamma$ is not required for our main result, Theorem \ref{Thm_main}.} Then the following points are equivalent:
\begin{enumerate}[(i)]
\item There exists a coordinate transformation $x^i\mapsto y^{\alpha}$ with Jacobian $J \in W^{m+1,p}(\Omega)$ such that $\Gamma^\alpha_{\beta\gamma}\in W^{m+1,p}(\Omega)$ in $y$-coordinates.
\item There exists a symmetric $(1,2)$-tensor $\Gammati \in W^{m+1,p}(\Omega)$ and a matrix field $J \in W^{m+1,p}(\Omega)$ which solve
\begin{eqnarray} 
J^{-1}dJ &=&\Gamma-\tilde{\Gamma}, \label{Riemann_flat_J-form_prelim} \\
 J^\alpha_{i,j} - J^\alpha_{j,i} &=& 0. \label{J_integrability_prelim} \label{J_integrability}
\end{eqnarray}
\item There exists a symmetric $(1,2)$ tensor $\Gammati \in W^{m+1,p}(\Omega)$ such that $\Gamma-\Gammati$ is Riemann-flat,
\beq \label{Riemann_flat_condition}
{\rm Riem} (\Gamma-\tilde{\Gamma}) =0.
\eeq   
\item There exists a symmetric $(1,2)$ tensor $\Gammati \in W^{m+1,p}(\Omega)$ which, when viewed as a matrix valued $1$-form in $x$-coordinates,  solves
\beq \label{Riemann_flat_curl-form_prelim}
d\tilde{\Gamma} = d\Gamma + \big(\Gamma - \tilde{\Gamma}\big) \wedge \big(\Gamma - \tilde{\Gamma}\big) .
\eeq
\end{enumerate}
\end{Thm}

\Proof
The equivalence of (i) and (ii) follows from \eqref{prelim_connection_transfo_1} - \eqref{prelim_connection_transfo_2}, where \eqref{J_integrability_prelim} is the Frobenius integrability condition.   That (ii) implies (iii) follows because \eqref{prelim_connection_transfo_2} implies the Riemann-flat condition \eqref{Riemann_flat_condition}.   The equivalence of (iii) and (iv) follows from Lemma \ref{Lemma_Riemann_2-form}.    Finally, the implication (iii) to (i) is established in \cite{ReintjesTemple_geo} in the case of the lower regularity class $\Gamma\in L^{\infty}$, $\tilde{\Gamma},J\in C^{0,1}$.   The more regular case of $\Gamma\in W^{m,p}$, $\tilde{\Gamma},J\in W^{m+1,p}$ here,  follows by a similar argument using compactness in $W^{m,p}$ of the unit ball in $W^{m+1,p}$, in place of the Arzela-Ascoli theorem.  (Details omitted.\footnote{In fact, although the Riemann-flat condition motivates this paper, the formal proofs only use the straightforward implication (i) implies (iii), that if one can smooth the connection, then the Riemann flat condition holds.}) 
\QED

In Sections \ref{Sec_equiv_warm-up} - \ref{Sec_equiv_main} we derive a pair of nonlinear Poisson equations equivalent to the Riemann-flat condition in the form (\ref{Riemann_flat_J-form_prelim}), such that it closes up in the unknowns $(J,\Gammati)$, with regularity in each term formally consistent with $\Gamma\in W^{m,p}$, but $\tilde{\Gamma},J\in W^{m+1,p}$.  This accomplishes the first step in our program to apply elliptic regularity results to solve the problem of regularity singularities.    To start, observe that equations  \eqref{Riemann_flat_J-form_prelim} - \eqref{J_integrability_prelim} are under-determined for unknowns $(J,\Gammati)$. On the other hand, \eqref{Riemann_flat_curl-form_prelim} is a system of equations for $\tilde{\Gamma}$ alone which is consistent with $\Gammati$ being one degree more regular than $\Gamma$, $\tilde{\Gamma}\in W^{m+1,p}$, but a necessary condition to solve them is that the exterior derivative of the right hand side must vanish.   The latter imposes additional conditions on $\tilde{\Gamma}$ that must be satisfied.   The objective of Sections \ref{Sec_equiv_warm-up} - \ref{Sec_equiv_main} is to derive equations \eqref{eqn1} - \eqref{eqn4} from \eqref{Riemann_flat_J-form_prelim} - \eqref{J_integrability_prelim}, a system of elliptic PDE's which closes up in $(J,\Gammati)$, and prove that finding solutions of this PDE is equivalent to solving the Riemann-flat condition \eqref{Riemann_flat_J-form_prelim} - \eqref{J_integrability_prelim}.

\section{Cauchy Riemann systems and Poisson equations}   \label{Sec_Cauchy-Riem}

In this section we get started by briefly reviewing the classical equivalence between Poisson equations and Cauchy Riemann type equations for matrix valued differential forms at the level of smoothness we are dealing with.  This is the starting point for our derivation of the elliptic system \eqref{eqn1} - \eqref{eqn4} in Sections \ref{Sec_equiv_warm-up} and \ref{Sec_equiv_main}.   The Riemann-flat condition is stated in terms of exterior derivatives, and we apply the ideas in this section to replace the $J$ equation, which as a first order equations is formally unsolvable, into a second order Poisson equation which is solvable.  The starting point is the following classical result for scalar valued differential forms, c.f. \cite{Dac}.  (We prove a generalization of this in Lemma \ref{Lemma_basic_smooth} below.) \\

\noindent {\bf Theorem:}  
{\it Assume $f$ is a smooth $(k+1)$-form and $g$ is a smooth $(k-1)$-form such that $df=0$ and $\delta g=0$. Then a $k$-form $u$ solves
\beq \label{firstorder}
du=f \hspace{1cm} \text{and} \hspace{1cm} \delta u = g
\eeq
if and only if $u$ solves 
\beq \label{secondorder}
\Delta u = \delta f + d g
\eeq
with boundary data $du=f$ and $\delta u = g$ on $\partial\Omega$. } \\

To introduce some ideas underlying Theorem \ref{Thm_main}, we now state and record the proof of a version of the classical result regarding the equivalence of \eqref{firstorder} - \eqref{secondorder}, which applies to solutions of nonlinear PDE's  involving matrix valued differential forms which more closely model \eqref{eqn1} - \eqref{eqn4}. For this, assume $f$ maps $k$-forms to $(k+1)$-forms and $g$ maps $k$-forms to $(k-1)$-forms, let $\Lambda_k^{m,p}(\Omega)$ denote the space of matrix-valued $k$-forms with components in $W^{m,p}(\Omega)$, and assume that
\begin{eqnarray} \label{f_g_spaces}
&f:& \Lambda_k^{m+1,p}(\Omega) \ \longrightarrow \  \Lambda_{k+1}^{m,p}(\Omega), \cr
&g:& \Lambda_k^{m+1,p}(\Omega) \ \longrightarrow \  \Lambda_{k-1}^{m,p}(\Omega). 
\end{eqnarray}
The loss and gain of derivatives in $f$ and $g$ are introduced to model the right hand side of  \eqref{eqn1} - \eqref{eqn4}.

\begin{Lemma} \label{Lemma_basic_smooth}              
Assume $f$ and $g$ as in \eqref{f_g_spaces}, and assume $m\geq2$, $1<p<\infty$, such that $d\big(f(w)\big)=0$ and $\delta \big(g(w)\big)=0$ for any $w\in \Lambda_k^{m+1,p}(\Omega)$. Then $u\in \Lambda_k^{m+1,p}(\Omega)$ solves 
\beq \label{Cauchy_Riem_smooth}
du=f(u) \hspace{1cm} \text{and} \hspace{1cm} \delta u = g(u),
\eeq
if and only if $u$ solves 
\beq \label{Laplacian_gen_smooth}
\Delta u = \delta \big(f(u) \big) + d \big( g(u) \big)
\eeq
with boundary data 
\beq \label{Cauchy_Riem_bd_smooth}
du=f \hspace{1cm} \text{and} \hspace{1cm} \delta u = g \hspace{1cm} \text{on} \ \ \partial \Omega.
\eeq
\end{Lemma}

\Proof
To prove that \eqref{Cauchy_Riem_smooth} implies \eqref{Laplacian_gen_smooth}, recall that $\Delta \equiv d \delta + \delta d$ by \eqref{def_Laplacian}.   Taking $\delta$ of $du=f(u)$ and $d$ of $\delta u =g(u)$ and adding the resulting equations, gives \eqref{Laplacian_gen_smooth}, and restricting \eqref{Cauchy_Riem_smooth} to $\partial \Omega$ gives \eqref{Cauchy_Riem_bd_smooth}. This proves the forward implications.

For the backward implication, assume \eqref{Laplacian_gen_smooth} and \eqref{Cauchy_Riem_bd_smooth}. To show that $u$ solves $du=f(u)$,  take the exterior derivative $d$ of the Poisson equation \eqref{Laplacian_gen_smooth}. Observing that $\Delta \equiv d \delta + \delta d$ commutes with $d$ (and $\delta$) and using $d^2=0$ and $df(u)=0$, we obtain
\beq \label{techeqn_basic_smooth_1}  
\Delta \big( du-f(u) \big) =0.
\eeq
Thus each component of $du-f(u)$ is a harmonic function in $\Omega$. Moreover, by \eqref{Cauchy_Riem_bd_smooth}, each component of $du-f(u)$ vanishes on the boundary, implying $du-f(u)=0$ in $\Omega$, thereby establishing the first equation in \eqref{Cauchy_Riem_smooth}.
Similarly, taking $\delta$ of \eqref{Laplacian_gen_smooth}, using $\delta^2=0$ and $\delta g(u)=0$, we find
\beq \label{techeqn_basic_smooth_2} 
\Delta \big( \delta u-g(u) \big) =0,
\eeq
which combined with boundary data \eqref{Cauchy_Riem_bd_smooth} implies $\delta u-g(u)=0$ in $\Omega$, so the second equation in \eqref{Cauchy_Riem_smooth} also holds.   This proves the backward implication.                
\QED
The above theorem and proof are correct at the level of classical derivates, but for the $A$ equation in system \eqref{eqn1} - \eqref{eqn4} we need to see that Lemma \ref{Lemma_basic_smooth} holds for solutions one degree less regular.   We state this as a lemma:

\begin{Lemma} \label{Lemma_basic_low-reg}              
Lemma \ref{Lemma_basic_smooth} is also true for $m\geq1$, $1<p<\infty$.
\end{Lemma}

\Proof  
The forward implication follows as in Lemma \ref{Lemma_basic_smooth} because the boundary data makes sense in $L^p$ by the trace theorem, \cite{Evans}.

For the backward implication at the lower regularity $m=1$, we must take derivatives in a distributional sense.   For this, take the $L^2$ inner product on matrix valued forms to be 
\beq \label{Def_L2-inner-product}
\langle \cdot , \cdot  \rangle_{L^2} \equiv \int_\Omega \tr \big(\langle \cdot\; ; \cdot \rangle \big),
\eeq
where $\tr(\cdot)$ denotes the trace of a matrix and $\langle \cdot\; ; \cdot \rangle $ is the matrix valued inner product defined in \eqref{def_inner-product}.  Using the definition of Hodge star \eqref{def_Hodge}, the product rule \eqref{Leibniz_rule} for matrix value forms, and Stokes Theorem, its easy to see that the standard integration by parts formula for $k$-forms extends to matrix valued forms,            
\beq \label{integration_by_parts}
\langle dw, v \rangle_{L^2} + \langle w, \delta v \rangle_{L^2} =0 ,
\eeq
where $w$ is a matrix valued $k$-form and $v$ a matrix valued $k+1$-form, both differentiable and at least one of them vanishing on $\partial\Omega$, (c.f. \cite[Theorem 1.11]{Dac}).

Now assume $\Delta u=\delta f+d g$ holds in $\Omega$, $du=f$, $\delta u=g$ on $\partial\Omega$, and $u\in W^{2,p}$.  We show $du=f(u)$ holds in the $L^p$ sense.   By Riesz representation, it suffices to show that
\beq  \label{techeqn0_Lemma_basic_low-reg}
\langle (du-f), \phi \rangle_{L^2} =0,
\eeq
for all $\phi \in L^{p^*}(\Omega)$, where $\frac1{p^*} + \frac1{p} =1$. Since the Laplacian is invertible, there exist a $\psi \in W^{2,p^*}(\Omega)$ such that $\Delta \psi = \phi$, and $\psi=0$ on $\partial\Omega$. Since by assumption, $du-f(u)=0$ on $\partial\Omega$, we can apply the integration by parts formula \eqref{integration_by_parts} and compute
\begin{eqnarray} \label{techeqn1_Lemma_basic_low-reg}
\langle (du-f), \phi \rangle_{L^2} 
&=& \langle (du-f), \Delta \psi \rangle_{L^2}  \cr
&=& -\langle \delta (du-f), \delta \psi \rangle_{L^2}  - \langle d(du-f), d \psi \rangle_{L^2} \cr
&=& -\langle \delta (du-f), \delta \psi \rangle_{L^2},
\end{eqnarray}
where in the last equality we use $d^2u= df=0$.  Since $\delta^2=0$ and $\delta u-g(u)=0$  on $\partial\Omega$, we have
\beq \nonumber
0 = \langle (\delta u-g ), \delta^2 \psi \rangle_{L^2} = - \langle d(\delta u-g ), \delta \psi \rangle_{L^2}.
\eeq
Adding this to \eqref{techeqn1_Lemma_basic_low-reg}, gives
\begin{eqnarray} \nonumber
\langle (du-f), \phi \rangle_{L^2} 
&=& -\langle \delta (du-f), \delta \psi \rangle_{L^2} - \langle d(\delta u-g ), \delta \psi \rangle_{L^2} \cr 
&=&  \langle (\Delta u- \delta f - dg , \delta \psi \rangle_{L^2} \ = \ 0,
\end{eqnarray}
which proves $du-f(u)=0$ in $\Omega$. A similar reasoning proves that $\delta u =g(u)$ holds as well.  This completes the proof.
\QED

\section{A first equivalence of the Riemann-flat condition to an elliptic system} \label{Sec_equiv_warm-up}

In this section we derive a system of nonlinear Poisson equations equivalent to the Riemann-flat condition in the case  $\Gamma$ and $\Riem(\Gamma)\in W^{m,p}(\Omega)$, $m\geq 1$, $p>n$.  For $m\geq 1$, solutions of the RT-equations are regular enough to impose boundary conditions, (c.f. Lemma \ref{Lemma_basic_low-reg}), and $p>n$ guarantees $W^{m,p}$ is closed under taking products.\footnote{Since the nonlinearities in the equations involve products of functions in $L^p$, (and more generally in $W^{m,p}$), and products of $L^p$ functions are not generally in $L^p$, we assume $p>n$ so the Sobelev embedding implies $L^p$ functions are H\"older continuous. Then we can estimate $\|fg\|_{p}\leq\|f\|_{L^\infty}\|g\|_{L^p}$ for $f,g\in W^{1,p}$, and similarly for $f,g$ in $W^{m,p}$.} Assuming $\Riem(\Gamma) \in W^{m,p}(\Omega)$ is equivalent to assuming $d\Gamma \in W^{m,p}(\Omega)$, so only $\delta \Gamma$ is free to be one level less smooth than $\Gamma$ and $d\Gamma$.  

To begin, recall that by Theorem \ref{Thm_Prelim} the Riemann-flat condition $Riem(\Gamma-\Gammati)=0$ can be written equivalently as
\begin{eqnarray} 
d\tilde{\Gamma} = d\Gamma + \big(\Gamma - \tilde{\Gamma}\big) \wedge \big(\Gamma - \tilde{\Gamma}\big).  \label{Riemann_flat_curl-form}
\end{eqnarray}
Condition \eqref{Riemann_flat_curl-form} is an equation on $\Gammati$ alone (not involving $J$), but it is not a solvable equation for $d\Gammati$, in part because we need to impose the consistency condition that $d$ of the right hand side of (\ref{Riemann_flat_curl-form}) be zero.
To complete (\ref{Riemann_flat_curl-form}) to a solvable system of equations, we look to couple (\ref{Riemann_flat_curl-form}) to an equation for the Jacobian $J$.
To construct such a system, we start with the equivalent expression of the Riemann-flat condition in terms of $J$, given by Lemma \ref{Thm_Prelim}, (c.f. \eqref{Riemann_flat_curl-form_prelim} and \eqref{Riemann_flat_J-form_prelim}),
\begin{eqnarray} 
J^{-1} dJ = \big( \Gamma-\tilde{\Gamma} \big). \label{Riemann_flat_J-form} 
\end{eqnarray}
The following lemma explains why the exterior derivative of the right hand side of (\ref{Riemann_flat_curl-form}) vanishes when coupled to \eqref{Riemann_flat_J-form}, since (\ref{Riemann_flat_curl-form}) together with equation (\ref{wedge}) of the lemma implies that
\beq \label{Riemann_d-identity}
 \big(\Gamma - \tilde{\Gamma}\big) \wedge \big(\Gamma - \tilde{\Gamma}\big) = - d\big(J^{-1} dJ\big).
\eeq        

\begin{Lemma} \label{Lemma_Riemann_d-form}                  
Any matrix valued $0$-form $J\in W^{2,p}(\Omega)$ satisfies               
\begin{eqnarray} 
d\big(J^{-1} dJ\big) &=& d(J^{-1}) \wedge dJ  \label{wedge_simple}  \\
&=& -\big(J^{-1} dJ\big)\wedge\big(J^{-1} dJ\big) . \label{wedge}
\end{eqnarray}
\end{Lemma}  

\Proof
Since the exterior derivative defined in \eqref{def_exterior_deriv} acts component-wise on matrix valued forms, it follows that $d^2J=0$. By the Leibniz rule for $k$-forms \eqref{Leibniz_rule}, we obtain that              
\begin{eqnarray} \nonumber
d\big(J^{-1} dJ\big) &=& d\big(J^{-1}\big)\wedge dJ+J^{-1}d^2J \cr 
&=& d(J^{-1}) \wedge dJ,
\end{eqnarray}
which gives \eqref{wedge_simple}. Moreover, using the Leibniz rule to compute
\begin{eqnarray} \nonumber
d(J^{-1}J) 
&=& \partial_i (J^{-1} J) dx^i  \cr
&=& \partial_i J^{-1} dx^i J + J^{-1} \partial_i J dx^i  \cr
&=& d(J^{-1})J+J^{-1}dJ,
\end{eqnarray}
we conclude from $d(J^{-1}J)=0$ that
$$
d\big(J^{-1}\big)= -J^{-1}\mm dJ \mm J^{-1}.
$$
The above identity now yields              
\begin{eqnarray} \nonumber
d(J^{-1}) \wedge dJ
= -J^{-1}dJJ^{-1}\wedge dJ,
\end{eqnarray}
and since matrices commute with basis elements of $k$-forms we have
\begin{eqnarray} \nonumber
d(J^{-1}) \wedge dJ
= -J^{-1}dJ\wedge J^{-1}dJ,
\end{eqnarray} 
which proves \eqref{wedge}.
\QED

We now derive a set of equations in $(\tilde{\Gamma},J)$ which is consistent and closes.  For the $\tilde{\Gamma}$ equations, in light of \eqref{Riemann_d-identity}, we take the Riemann-flat condition \eqref{Riemann_flat_condition} as
\beq \label{Riemann_flat_d-form}
d\tilde{\Gamma} = d \Gamma - d\big(J^{-1} dJ\big).
\eeq
The right hand side is consistent with the left hand side since both are exterior derivatives. (Note, taking the exterior derivative of (\ref{Riemann_flat_J-form}) gives \eqref{Riemann_flat_d-form}, thereby showing how information in (\ref{Riemann_flat_curl-form}) is encoded in (\ref{Riemann_flat_J-form}).)  Motivated by the fact that only $d\Gammati$ appears in the curvature, we allow $\delta\tilde{\Gamma}$ to be a free function, and set
\beq \label{delta-eqn-h}
\delta \tilde{\Gamma} = h,
\eeq
where $h\in W^{1,p}$ is an arbitrary matrix valued $0$-form. The freedom in choosing $h$ reflects the freedom in choosing smooth coordinate transformations to maintain the smoothness of a spacetime connection.

For fixed function $J$, one could solve \eqref{Riemann_flat_d-form} - \eqref{delta-eqn-h} for $\tilde{\Gamma}$ by the existence theory in \cite{Dac}, (the Poincar\'e Lemma), since the consistency condition is that the exterior derivative of the right hand side of \eqref{Riemann_flat_d-form} vanishes, and $\delta h=0$ for matrix valued $0$-forms. Alternatively, adding $\delta$ of \eqref{Riemann_flat_d-form} and $d$ of \eqref{delta-eqn-h} produces the second order Poisson equation
\beq
\Delta \Gammati = \delta d \big( \Gamma - J^{-1} dJ \big) + dh . \label{Gammati_Poisson}
\eeq
By Lemma \ref{Lemma_basic_smooth}, it follows that any solution of \eqref{Gammati_Poisson} which satisfies \eqref{Riemann_flat_d-form} - \eqref{delta-eqn-h} on $\partial\Omega$, is also a solution of the Cauchy-Riemann system \eqref{Riemann_flat_d-form} - \eqref{delta-eqn-h} in $\Omega$.

The problem of deriving the $J$ equation is not so simple. It turns out we \emph{need} a second order equation, since the consistency condition that the right hand side of the first order equation \eqref{Riemann_flat_J-form} have a vanishing exterior derivative, leads to circularity. To see this, we introduce the following lemma. 

\begin{Lemma} \label{Lemma_basic_J_problem}
Assume \eqref{Riemann_flat_d-form} holds for $J,\Gammati \in W^{m+1,p}(\Omega)$ and $\Gamma \in W^{m,p}(\Omega)$ for $m\geq 1$, then  
\beq \label{basic_J_problem}
d\Big( J\mm(\Gamma- \tilde{\Gamma})\Big) = dJ \wedge \Big( (\Gamma- \tilde{\Gamma}) - J^{-1} dJ \Big).
\eeq                      
\end{Lemma}

\Proof
A straightforward computation using the Leibniz rule for $k$-forms \eqref{Leibniz_rule} gives
\begin{eqnarray}
d\big( J\mm(\Gamma- \tilde{\Gamma})\big) 
&=& dJ \wedge (\Gamma- \tilde{\Gamma}) + J \mm (d\Gamma- d\tilde{\Gamma}) \cr
&=& dJ \wedge (\Gamma- \tilde{\Gamma}) + J \mm d(J^{-1} dJ),
\end{eqnarray}
where we used \eqref{Riemann_flat_d-form} for the last equality, and substituting \eqref{wedge} for $d(J^{-1} dJ)$ yields 
\beq \nonumber
d\big( J\mm(\Gamma- \tilde{\Gamma})\big) = dJ \wedge (\Gamma- \tilde{\Gamma}) - dJ \wedge J^{-1} dJ,
\eeq
which is the sought after equation \eqref{basic_J_problem}.
\QED

To see the circularity in the first order equation for $J$, note that one can solve the Riemann-flat condition \eqref{Riemann_flat_J-form} for $J$ only under the consistency condition that the exterior derivative of $J$ times its right hand side should vanish. By \eqref{basic_J_problem}, the exterior derivative of the right hand side vanishes if either either $dJ=0$, (in which case $\Gamma=\tilde{\Gamma}$, and $\tilde{\Gamma}$ does not have the regularity we seek), or if \eqref{Riemann_flat_J-form} holds, which just reproduces the equation for $J$ we started with, which is circular; or else the right hand side of (\ref{basic_J_problem}) produces a nonlinear PDE in $J$ more complicated than the one we started with. 

Thus, differently from the case of $\tilde{\Gamma}$,  we need a second order equation in $J$ in order to obtain a solvable PDE.   The second order equation for $J$ obtained from \eqref{Riemann_flat_J-form} is again a non-linear Poisson equation which does not require the constraint that the right hand side of  (\ref{basic_J_problem}) should vanish.  To obtain this, again use $\Delta= \delta d +d\delta$, and note that for the $0$-forms $J$ we have $\delta J=0$, so that $\Delta J=\delta d J$.   Then taking $\delta$ of equation \eqref{Riemann_flat_J-form},  we obtain 
\beq \label{Laplacian_J_derivation_totalderiv}
\Delta J = \delta \big( J \mm ( \Gamma - \tilde{\Gamma} ) \big) .
\eeq
Applying \eqref{colon} gives 
$$
\delta (J\mm \Gammati) =  J \mm \delta \Gammati + \langle dJ ; \Gammati \rangle.
$$ 
Thus, replacing $\delta \Gammati = h$ by \eqref{delta-eqn-h} yields equation   \eqref{Laplacian_J_derivation_totalderiv} in its final form, 
\beq \label{Laplacian_J_derivation}
\Delta J = \delta \big( J \mm \Gamma ) -  J \mm h - \langle d J ; \tilde{\Gamma} \rangle,
\eeq
where again $h$ is a free function.   In contrast to the first order equation \eqref{Riemann_flat_J-form}, solving the nonlinear Poisson equations \eqref{Laplacian_J_derivation} allows for more general boundary data and does not require the right hand side to have a vanishing exterior derivative. 

To summarize, every solution $(J,\tilde{\Gamma})$ of the Riemann-flat condition \eqref{Riemann_flat_J-form} also solves the second order equations \eqref{Gammati_Poisson} and \eqref{Laplacian_J_derivation}. In the next theorem we prove equivalence of \eqref{Gammati_Poisson} and \eqref{Laplacian_J_derivation} with the Riemann-flat condition \eqref{Riemann_flat_J-form}, in the sense that a solution $(J,\tilde{\Gamma})$ of the Poisson system \eqref{Gammati_Poisson} and \eqref{Laplacian_J_derivation} gives rise to a solution of the original Riemann-flat condition \eqref{Riemann_flat_J-form} after suitable modification of $\tilde{\Gamma}$. Remarkably, in contrast to Lemma \ref{Lemma_basic_smooth}, the second order system \eqref{Gammati_Poisson} and \eqref{Laplacian_J_derivation} generates solutions of the first order system without requiring any boundary conditions.

\begin{Thm} \label{Thm0}
Let $\Gamma, d\Gamma \in W^{m,p}(\Omega)$, for $m\geq 1$, $p>n$.  Then if $(J,\tilde{\Gamma})$ solves the Riemann-flat condition \eqref{Riemann_flat_J-form} for a matrix valued $0$-form $J \in W^{m+1,p}(\Omega)$ and a matrix valued $1$-form $\tilde{\Gamma} \in W^{m+1,p}(\Omega)$, then $(J,\tilde{\Gamma})$ solves
\begin{eqnarray} 
\Delta \Gammati &=& \delta d \Gamma - \delta \big( d(J^{-1}) \wedge dJ \big) + dh , \label{Laplacian_Gammati} \\
\Delta J &=& \delta ( J \mm \Gamma ) -  J \mm h - \langle d J ; \tilde{\Gamma}\rangle , \label{Laplacian_J}
\end{eqnarray}
for $h \equiv \delta \Gammati \in W^{m,p}(\Omega)$. 
Conversely, if $(J,\tilde{\Gamma})\in W^{m+1,p}(\Omega)$ solves \eqref{Laplacian_Gammati}-\eqref{Laplacian_J} for some matrix-valued $0$-form $h\in W^{m,p}(\Omega)$, then, defining
\begin{eqnarray} \label{defhatGamma}    
\tilde{\Gamma}' \equiv \Gamma -J^{-1}dJ ,
\end{eqnarray}
the modified pair $(J,\tilde{\Gamma}')$ solves the Riemann-flat condition \eqref{Riemann_flat_J-form} with  $\tilde{\Gamma}' \in W^{m+1,p}(\Omega')$, (the regularity required by the equivalences of Theorem \ref{Thm_Prelim}), on any compactly contained open set $\Omega' \subset \subset \Omega$.
\end{Thm}

\Proof
For the forward implication, assume that $J$ and $\Gammati$ satisfy the Riemann-flat condition \eqref{Riemann_flat_J-form} in $\Omega$. Taking the exterior derivative $d$ of \eqref{Riemann_flat_J-form} implies  \eqref{Riemann_flat_d-form}, while \eqref{delta-eqn-h} follows by definition of $h$. Adding now $\delta$ of \eqref{Riemann_flat_d-form} and $d$ of \eqref{delta-eqn-h} gives 
\beq \nonumber
\Delta \Gammati = \delta d \Gamma -  \delta d \big( J^{-1} dJ \big) + dh,
\eeq
and applying the identity $d ( J^{-1} dJ) = d(J^{-1}) \wedge dJ$ of Lemma \ref{Lemma_Riemann_d-form} yields the sought after Poisson equation \eqref{Laplacian_Gammati}.\footnote{Let us remark that we could have established the equivalence of Theorem \ref{Thm0} for \eqref{Laplacian_J} together with the first order system \eqref{Riemann_flat_d-form} - \eqref{delta-eqn-h} for $\Gammati$. However, system \eqref{Laplacian_Gammati} - \eqref{Laplacian_J} is preferable, since the existence theory for the first order system \eqref{Riemann_flat_d-form} - \eqref{delta-eqn-h} in \cite{Dac} is more delicate than for the Poisson-type equation \eqref{Laplacian_Gammati}, c.f. \cite{ReintjesTemple_ell2}.} The  argument leading to \eqref{Laplacian_J_derivation} shows that any solution of \eqref{Riemann_flat_J-form} also solves the Poisson equation \eqref{Laplacian_J} for $J$. This proves the forward implication. 

To prove the backward implication, assume $(J,\tilde{\Gamma})\in W^{m+1,p}(\Omega)$ solves \eqref{Laplacian_Gammati} - \eqref{Laplacian_J} for some matrix-valued $0$-form $h\in W^{m,p}(\Omega)$. 
Then, by definition \eqref{defhatGamma}, $(J,\tilde{\Gamma}')$ solves the Riemann-flat condition \eqref{Riemann_flat_J-form} and $\Gammati' \in W^{m,p}(\Omega)$, since $J^{-1} dJ \in W^{m,p}(\Omega)$ and $\Gamma \in W^{m,p}(\Omega)$. The nontrivial part of the proof is to show that $\Gammati'$ is one degree more regular than the terms $J^{-1} dJ$ and $\Gamma$ on the right hand side of \eqref{defhatGamma}, that is, to show that $\tilde{\Gamma}' \in W^{m+1,p}$. For this, we first show that 
\beq \label{Poisson_Gammati'_reg}
\Delta \Gammati' \in W^{m-1,p}(\Omega)     ,
\eeq
so that standard estimates of elliptic regularity theory imply the sought after regularity $\Gammati'\in W^{m+1,p}(\Omega')$ for any open set $\Omega' \subset \subset \Omega$.

We now establish \eqref{Poisson_Gammati'_reg}. For this, we first use \eqref{wedge_simple} of Lemma \ref{Lemma_Riemann_d-form} to write equation \eqref{Laplacian_Gammati} in the equivalent form
\beq \nonumber
\Delta \Gammati = \delta d \big( \Gamma -  J^{-1} dJ \big) + dh ,
\eeq
so that substituting $\tilde{\Gamma}' \equiv \Gamma -J^{-1}dJ$ gives  
\beq \label{Proof_Thm0_eqn1}
\delta d \tilde{\Gamma}' = \Delta \Gammati - dh .
\eeq
To determine the second term on the right hand side of $\Delta \Gammati' = \delta d \Gammati' +d \delta \Gammati'$, we compute
\begin{eqnarray}  \label{Proof_Thm0_eqn2}
\delta \tilde{\Gamma}' 
&\overset{\eqref{defhatGamma}}{=}& \delta \Gamma - \delta \big( J^{-1} dJ \big) \cr
&\overset{\eqref{colon}}{=}& \delta \Gamma  -  \langle d (J^{-1}) ;  dJ \rangle   -  J^{-1} \delta d J   \cr
&=& \delta \Gamma  -  \langle d (J^{-1}) ;  dJ \rangle   -  J^{-1} \Delta J,  
\end{eqnarray}
where the last equality follows since $\Delta J = \delta d J$ for $0$-forms. Substituting \eqref{Laplacian_J} gives
\begin{eqnarray}  \nonumber
\delta \tilde{\Gamma}' 
&=& \delta \Gamma  -  \langle d (J^{-1}) ;  dJ \rangle   -  J^{-1} \Big( \delta ( J \mm \Gamma ) -  J \mm h - \langle d J ; \tilde{\Gamma}\rangle \Big)   ,
\end{eqnarray}
and since $\delta ( J \mm \Gamma ) = J \delta \Gamma + \langle d J ; \Gamma \rangle$ by equation \eqref{colon} we obtain that
\begin{eqnarray} \label{Proof_Thm0_eqn3}
\delta \tilde{\Gamma}' 
&=&  -  \langle d (J^{-1}) ;  dJ \rangle  + h -  J^{-1} \langle d J ; \Gamma - \tilde{\Gamma} \rangle   ,
\end{eqnarray}
where the terms containing $\delta \Gamma$ canceled, resulting in a gain of one derivative.
Taking now the exterior derivative $d$ of \eqref{Proof_Thm0_eqn3} and adding the resulting equation to \eqref{Proof_Thm0_eqn1} results in
\beq \label{Poisson_Gammati'}
\Delta \tilde{\Gamma}'  = \Delta \Gammati   -  d \Big( \langle d J^{-1} ;  dJ \rangle  + J^{-1}\langle d J ; \Gamma - \tilde{\Gamma} \rangle \Big)
\eeq
Since $(J, \Gammati) \in W^{m+1,p}(\Omega)$ and since by Lemma \ref{Lemma_Leibnitz-rule} products of functions in $W^{m,p}(\Omega)$ are again in $W^{m,p}(\Omega)$ for $m\geq1$, $p>n$, it follows that the right hand side of \eqref{Poisson_Gammati'} is in $W^{m-1,p}(\Omega)$ which proves \eqref{Poisson_Gammati'_reg}.

To complete the proof, we apply the elliptic estimate \eqref{elliptic_estimate_Lp_Lemma} of Lemma \ref{Lemma_elliptic_estimate} component-wise to \eqref{Poisson_Gammati'}. In more detail, in each fixed component the Poisson equation \eqref{Poisson_Gammati'} is of the form
\beq \nonumber
\Delta u = f
\eeq
for some scalar valued functions $f \in \  W^{m-1,p}(\Omega)$ and $u \in W^{m,p}(\Omega)$. Then, by elliptic estimate \eqref{elliptic_estimate_Lp_Lemma} of Lemma \ref{Lemma_elliptic_estimate}, there exists for each $\Omega' \subset \subset \Omega$ a constant $C>0$ (depending only on $\Omega'$, $\Omega$, $m,n,p$), such that
\beq \label{elliptic_estimate_Lp_2}
\| u \|_{W^{m+1,p}(\Omega')} \leq C \big( \| f \|_{W^{m-1,p}(\Omega)} + \| u \|_{W^{m-1,p}(\Omega)} \big).
\eeq
Thus, since $\Gammati' \in W^{m,p}(\Omega)$ and since the right hand side of \eqref{Poisson_Gammati'} is in $W^{m-1,p}(\Omega)$, estimate \eqref{elliptic_estimate_Lp_2} implies the sought after gain of one derivative, i.e. $\Gammati' \in W^{m+1,p}(\Omega')$ for any open set $\Omega' \subset \subset \Omega$.
\QED

In summary, (\ref{Riemann_flat_curl-form}) and (\ref{Riemann_flat_J-form}), being equivalent forms of the Riemann-flat condition, are not independent.  But reassured by the fact that the Riemann-flat condition is necessary and sufficient for metric smoothing, we apply the identity $\Delta=d\delta+\delta d$ to (\ref{Riemann_flat_curl-form}) and (\ref{Riemann_flat_J-form}) separately to construct two independent equations in $\Delta\Gammati$ and $\Delta J$. At the end, we use the ``gauge freedom'' in $\delta \Gammati = h$ to prove that solutions $(\Gammati,J,h)$ can always be transformed into solutions $(\Gammati',J,h')$, where $\Gammati'$ solves the Riemann-flat condition (and hence the RT-equations)  at the appropriate order of smoothness.   At the end, the authors find it remarkable that two equations, which at the start are equivalent, can be transformed into independent solvable equations (the RT-equations), and yet the independent $J$ and $\Gammati$ solving the RT-equations can always be transformed into a $\Gammati'$ which solves \eqref{SmoothingCondition} (and hence the Riemann-flat condition) with the same $J$.

\section{Our main equivalence theorem} \label{Sec_equiv_main}

We now consider the problem of imposing \eqref{J_integrability_prelim}, that is, the condition that $J$ be a true Jacobian, integrable to coordinates. The goal of this section is to augment system (\ref{Laplacian_Gammati})-(\ref{Laplacian_J}) with a first order PDE on the free function $h$ in (\ref{Laplacian_Gammati})-(\ref{Laplacian_J}) to replace the integrability condition (\ref{J_integrability_prelim}). 
Assume again throughout that $\Gamma, {\rm Riem}(\Gamma) \in W^{m,p}(\Omega)$ for $m\geq 1$ and $p>n$. 

The key idea to augment system  \eqref{Laplacian_Gammati} - \eqref{Laplacian_J} with an additional equation for the free function $h$ which is equivalent to \eqref{J_integrability} expressed in terms of exterior derivatives. To accomplish this, note first that the integrability condition \eqref{J_integrability} is equivalent to
\beq \label{J_integrability_vec}
d\vec{J}= 0,
\eeq
since                
\beq \nonumber 
Curl(J)^\alpha \equiv \frac12 \big( J^\alpha_{i,j} - J^\alpha_{j,i} \big) dx^j \otimes dx^i = J^\alpha_{i,j} dx^j\wedge dx^i =  d(J^\alpha_i dx^i) \equiv  d\vec{J}^\alpha .
\eeq 
Now, to combine \eqref{J_integrability_vec} with the Poisson equation \eqref{Laplacian_J}, observe that 
\beq \label{laplacian}
\overrightarrow{\Delta J} = (\Delta J^\alpha_i) dx^i =  \Delta (J^\alpha_i dx^i) =\Delta \vec{J},
\eeq  
since $\Delta$ acts component-wise on matrix valued $k$-forms by \eqref{basics_Laplacian}. Thus, interpreting the Poisson equation \eqref{Laplacian_J} in a vector sense, applying \eqref{laplacian} and taking $d$ of the resulting equation \eqref{Laplacian_J}, we obtain
\beq \nonumber 
\Delta d\vec{J} = d \big(\overrightarrow{\delta ( J \mm \Gamma )}\big) - d\big(\overrightarrow{ J \mm h}\big) -  d\big(\overrightarrow{ \langle d J ; \tilde{\Gamma}\rangle }\big),
\eeq
where we used that $\Delta$ and $d$ commute. Therefore, if $J$ solves \eqref{J_integrability_vec} in addition to \eqref{Laplacian_J}, then $A\equiv J\mm h$ must satisfy the equation
\beq  \label{J_intergability_on_h}
d \vec{A} = d \big(\overrightarrow{\delta ( J \mm \Gamma )}\big) - d\big(\overrightarrow{\langle d J ; \tilde{\Gamma}\rangle }\big).
\eeq
The right hand side of \eqref{J_intergability_on_h} is a vector valued $2$-form and vanishes when taking its exterior derivative (since $d^2=0$) so that \eqref{J_intergability_on_h} is well-posed for $A$ given $J$ and $\Gammati$. Our next goal is to show the backward implication,  that \eqref{J_intergability_on_h} together with the Poisson equation \eqref{Laplacian_J} on $J$ imply \eqref{J_integrability_vec}. 

\begin{Lemma} \label{Lemma_integrability}
Let $\Gamma \in W^{m,p}(\Omega)$ for $p>n$ and $m\geq 1,$ and let $\tilde{\Gamma}\in W^{m+1,p}(\Omega)$, $J\in W^{m+1,p}(\Omega)$ and $A \in W^{m,p}(\Omega)$ be given. Assume $J$ solves
\beq 
\Delta J = \delta ( J \mm \Gamma ) - \langle d J ; \tilde{\Gamma}\rangle - A \label{Lemma_integr_PDE1},
\eeq
$($the Poisson equation \eqref{Laplacian_J} with  $h= J^{-1} A)$.  Then $J$ satisfies the Curl-free condition \eqref{J_integrability_vec}, if and only if $A$ solves
\begin{eqnarray} 
d \vec{A} &=& d \big(\overrightarrow{\delta ( J\mm \Gamma )}\big) - d\big(\overrightarrow{\langle d J ; \tilde{\Gamma}\rangle }\big) \label{Lemma_integr_PDE2}
\end{eqnarray}
and
\begin{eqnarray} \label{boundary}
d\vec{J}=0\ \ \ \text{on }\partial\Omega.
\end{eqnarray}            
\end{Lemma}

\Proof
For the forward implication, assume $J$ solves \eqref{J_integrability_vec}. Then $A\equiv J\mm h$ solves \eqref{Lemma_integr_PDE2} by the argument in \eqref{J_integrability_vec} through \eqref{J_intergability_on_h}. Moreover, (\ref{boundary}) follows upon restriction of \eqref{J_integrability_vec} to $\partial\Omega$, (using that derivatives of $J$ are H\"older continuous because $p>n$). This proves the forward implication.

For the backward implication, assume $A$ solves \eqref{Lemma_integr_PDE2} and \eqref{boundary}.   Now, consider \eqref{Lemma_integr_PDE1} as an equation on vector valued $1$-forms and assume for the beginning that $m \geq 2$. Then, taking $d$ of \eqref{Lemma_integr_PDE1}, we get
\beq \nonumber
\Delta \big( d \vec{J} \big) = d \big(\overrightarrow{\delta ( J \mm \Gamma )}\big) - d\big(\overrightarrow{\langle d J ; \tilde{\Gamma}\rangle }\big)  - d\vec{A},
\eeq  
so that \eqref{Lemma_integr_PDE2} implies
\beq \nonumber 
\Delta \big( d\vec{J} \big)=0.
\eeq 
Therefore, since $d\vec{J}$ is assumed to vanish on $\partial\Omega$ as a H\"older continuous function, we conclude that \eqref{J_integrability_vec} holds in $\Omega$. This establishes the backward implication for $m\geq 2$.

Consider now the case that $m=1$, then $\Delta J \in L^p(\Omega)$ and we need to take $d$ in a distributional sense. For this, we proceed as in Lemma \ref{Lemma_basic_low-reg}: By Riesz representation, it suffices to show that
\beq  \label{techeqn0_Lemma_integrability}
\langle d\vec{J}, \phi \rangle_{L^2} = 0,
\eeq
for all scalar valued $2$-forms $\phi \in L^{p^*}(\Omega)$, where $\frac1{p^*} + \frac1{p} =1$, and where $\langle\cdot , \cdot \rangle_{L^2}$ denotes the standard $L^2$ inner product on differential forms which we apply component-wise to vector-valued forms. For each such $\phi$, there exists a scalar valued $2$-form $\psi \in W^{2,p^*}(\Omega)$ such that $\Delta \psi = \phi$, and $\psi=0$ on $\partial\Omega$. Using the product rule \eqref{Def_L2-inner-product} we compute
\begin{eqnarray} \label{techeqn1_Lemma_integrability}
\langle d\vec{J}, \phi \rangle_{L^2} 
&= & \langle d\vec{J}, \Delta\psi \rangle_{L^2} \cr 
& =&  - \langle \delta d\vec{J}, \delta {\psi} \rangle_{L^2}  \cr
& = & - \langle \Delta\vec{J}, \delta {\psi} \rangle_{L^2},
\end{eqnarray}
where the last equality follows since
\beq \nonumber
\langle d \delta \vec{J}, \delta {\psi} \rangle_{L^2}  = \langle \delta \vec{J}, \delta^2 {\psi} \rangle_{L^2} =0.
\eeq
Substituting now \eqref{Lemma_integr_PDE1} for $\Delta \vec{J} = \overrightarrow{\Delta J}$ in \eqref{techeqn0_Lemma_integrability}, we find
\begin{eqnarray} \label{techeqn2_Lemma_integrability}
\langle \Delta\vec{J}, \delta {\psi} \rangle_{L^2} 
&=& \Big\langle \overrightarrow{\delta ( J \mm \Gamma )} - \overrightarrow{\langle d J ; \tilde{\Gamma}\rangle} , \delta {\psi} \Big\rangle_{L^2} -  \langle \vec{A}, \delta {\psi} \rangle_{L^2} \cr
& = & \Big\langle \overrightarrow{\delta ( J \mm \Gamma )} - \overrightarrow{\langle d J ; \tilde{\Gamma}\rangle} , \delta {\psi} \Big\rangle_{L^2} + \langle d\vec{A}, {\psi} \rangle_{L^2} .
\end{eqnarray}
Substituting \eqref{Lemma_integr_PDE2} for $d \vec{A}$ and using the product rule one more time gives
\begin{eqnarray}
\langle d\vec{A}, {\psi} \rangle_{L^2} 
&=& \Big\langle d \big(\overrightarrow{\delta ( J\mm \Gamma )}\big) - d\big(\overrightarrow{\langle d J ; \tilde{\Gamma}\rangle }\big), {\psi} \Big\rangle_{L^2} \cr
&=& - \Big\langle \overrightarrow{\delta ( J\mm \Gamma )} - \overrightarrow{\langle d J ; \tilde{\Gamma}\rangle }, \delta {\psi} \Big\rangle_{L^2},
\end{eqnarray}
and substituting back into \eqref{techeqn2_Lemma_integrability}, a cancellation gives
\beq \nonumber
\langle d\vec{J}, \phi \rangle_{L^2}  = 0.
\eeq
This completes the proof.
\QED

Before we state our main theorem, we discuss the regularity of $A$. Since we seek $\Gammati \in W^{m+1,p}$ and $dh=d(J^{-1}A)$ is a source term on the right hand side of the Poisson equation \eqref{Laplacian_Gammati} for $\Gammati$, we need $A\in W^{m,p}$ (for $m\geq 1$) to be  consistent with $\Gammati \in W^{m+1,p}$. But this appears to contradict the fact that the first term on the right hand side of \eqref{Lemma_integr_PDE2} contains two derivatives on $\Gamma \in W^{m,p}$. Most remarkably, the consistency follows by our incoming assumption $d\Gamma \in W^{m,p}$ alone, in light of  identity \eqref{regularity-miracle} of Lemma \ref{Lemma_regularity-miracle}, 
\beq \nonumber
d \big(\overrightarrow{\delta ( J \mm \Gamma )}\big) 
= \overrightarrow{\text{div}} \big(dJ \wedge \Gamma\big) + \overrightarrow{\text{div}} \big( J\mm d\Gamma\big) ,
\eeq 
where $\overrightarrow{\text{div}}$ is defined in \eqref{Def_vec-div}. Therefore, since we assume $d\Gamma \in W^{m,p}(\Omega)$, we find that 
$$
d \big(\overrightarrow{\delta ( J \mm \Gamma )}\big) \in W^{m-1,p}(\Omega)
$$ 
and we conclude that the regularity of the right hand side of (\ref{Lemma_integr_PDE2}) is consistent with the regularity on the left hand side.

We now show that the existence of solutions $(J,\Gammati')$ of the Riemann-flat condition \eqref{Riemann_flat_J-form} together with the Curl-free condition \eqref{J_integrability_vec} is equivalent to the existence of solutions $(J,\Gammati,A)$ to a coupled system of non-linear elliptic equations, system \eqref{eqn1} - \eqref{eqn4}, and the equations are formally consistent at the levels of regularity we seek. This establishes Theorem \ref{Thm_main}.

\begin{Thm} \label{Thm1}
Let $\Gamma$ and ${\rm Riem}(\Gamma)$ be in $W^{m,p}(\Omega)$ for $p>n$ and $m\geq 1$. Then the following equivalence holds: \vspace{.15cm} \newline 
If there exists an invertible matrix-valued $0$-form $J \in W^{m+1,p}(\Omega)$ and a matrix-valued $1$-form $\tilde{\Gamma} \in W^{m+1,p}(\Omega)$ which solve 
\begin{eqnarray} \nonumber
&J^{-1} dJ = \Gamma - \Gammati ,& \cr
& d \vec{J}=0,\ &
\end{eqnarray}
c.f.  \eqref{Riemann_flat_J-form} and \eqref{J_integrability_vec}, then there exists $A\in W^{m,p}(\Omega)$ such that $(J,\Gammati,A)$ solve the elliptic system
\begin{eqnarray}
\Delta \Gammati &=& \delta d \Gamma - \delta \big( d(J^{-1})\wedge dJ \big) + d(J^{-1} A ) , \label{PDE_Gammati_Thm1} \label{PDE_Gammati-d_Thm1} \label{PDE_Gammati-delta_Thm1} \\
\Delta J &=& \delta ( J \mm \Gamma ) - \langle d J ; \tilde{\Gamma}\rangle - A \label{PDE_J_Thm1} , \\
d \vec{A} &=& \overrightarrow{\text{div}} \big(dJ \wedge \Gamma\big) + \overrightarrow{\text{div}} \big( J\, d\Gamma\big) - d\big(\overrightarrow{\langle d J ; \tilde{\Gamma}\rangle }\big)   \label{PDE_A_Thm1} \\
\delta \vec{A} &=& v \label{PDE_A-free_Thm1}
\end{eqnarray} 
in $\Omega$ with boundary data  
\begin{eqnarray}
d\vec{J}=0 \ \ \text{on} \ \partial\Omega,   \label{J_bdd_Thm1}
\end{eqnarray}  
where $v\in W^{m-1,p}(\Omega)$ is a vector valued $0$-form free to be chosen.   \vspace{.15cm} \newline
Conversely, if there exists $J \in W^{m+1,p}(\Omega)$ invertible, $\Gammati \in W^{m+1,p}(\Omega)$ and $A\in W^{m,p}(\Omega)$ solving \eqref{PDE_Gammati_Thm1} - \eqref{J_bdd_Thm1}, then there exists a $\Gammati' \in W^{m,p}(\Omega)$ such that for every $\Omega'$ compactly contained in $\Omega$ we have $\Gammati' \in W^{m+1,p}(\Omega')$ and $(J,\Gammati')$ solve \eqref{Riemann_flat_J-form} and \eqref{J_integrability_vec} in $\Omega'$.
\end{Thm}

\Proof
For the forward implication, assume there exists $\tilde{\Gamma} \in W^{m+1,p}(\Omega)$ and $J \in W^{m+1,p}(\Omega)$ which solve the Riemann-flat condition \eqref{Riemann_flat_J-form} together with the Curl-free condition \eqref{J_integrability_vec}. Theorem \ref{Thm0} implies that $J$ and $\Gammati$ solve \eqref{Laplacian_Gammati} - \eqref{Laplacian_J} for some $h \in W^{m,p}(\Omega)$, and setting $A=Jh$ it follows that $(J,\Gammati)$ solve \eqref{PDE_Gammati_Thm1} - \eqref{PDE_J_Thm1}.  Since $J$ satisfies \eqref{J_integrability_vec}, Lemma \ref{Lemma_integrability} implies that $A\in W^{m,p}(\Omega)$ solves \eqref{PDE_A_Thm1}. This proves the forward implication.

For the backward implication, assume $J \in W^{m+1,p}(\Omega)$, $\Gammati \in W^{m+1,p}(\Omega)$ and $A \in W^{m,p}(\Omega)$ solve the elliptic system \eqref{PDE_Gammati_Thm1} - \eqref{PDE_A_Thm1}, with $J$ invertible. Now, Theorem \ref{Thm0} implies that $J$ and $\Gammati' \equiv J^{-1} dJ - \Gamma$ solve the Riemann-flat condition \eqref{Riemann_flat_J-form} in each $\Omega'$ compactly contained in $\Omega$, and $\Gammati' \in W^{m+1,p}(\Omega')$ has the required regularity. Moreover, since \eqref{PDE_J_Thm1} and \eqref{PDE_A_Thm1} hold together with the boundary condition \eqref{J_bdd_Thm1}, Lemma \ref{Lemma_integrability} applies and yields that $J$ satisfies the integrability condition \eqref{J_integrability_vec} in $\Omega$ and therfore also in $\Omega'\subset\Omega$. This completes the proof.
\QED

Equations (\ref{PDE_Gammati_Thm1})-(\ref{PDE_A-free_Thm1}) are the fundamental equations of this paper, the RT-equations. Theorem \ref{Thm1} establishes our main theorem, Theorem \ref{Thm_main} of the Introduction, due to the equivalence of (i) and (ii) of Theorem \ref{Thm_Prelim}, which also holds true if $\Gamma$ and $\Gammati$ are not symmetric.  Again, authors prove an existence theory for (\ref{PDE_Gammati_Thm1})-(\ref{J_bdd_Thm1}) in \cite{ReintjesTemple_ell2}.

\section{An alternative equivalent elliptic system}   \label{Sec_equiv_system}

In this subsection, we prove the following proposition which shows that system \eqref{PDE_A_Thm1} can also be written equivalently as a system of coupled semi-linear Poisson equations, but to assign classical boundary data for $A$ we must assume one more order of smoothness than in Theorem \ref{Thm1}. 

\begin{Prop} \label{Cor2_Thm1}
Let $m\geq 2$ and assume that $\Gamma$ and $d\Gamma$ are both in $W^{m,p}(\Omega)$ for $p>n$. Let $(J,\tilde{\Gamma}) \in W^{m+1,p}(\Omega)$ solve \eqref{PDE_Gammati_Thm1} - \eqref{PDE_J_Thm1}, where $J$ is invertible. Then $A\in W^{m,p}(\Omega)$ solves \eqref{PDE_A_Thm1} in $\Omega$ if and only if $A$ solves
\begin{eqnarray} 
\Delta \vec{A}   &=& \delta \Big( \overrightarrow{\text{div}} \big(dJ \wedge \Gamma\big) + \overrightarrow{\text{div}} \big( J\, d\Gamma\big) - d\big(\overrightarrow{\langle d J ; \tilde{\Gamma}\rangle }\big) \Big) + d v,    \label{PDE_A_Cor2} 
\end{eqnarray}
in $\Omega$ with boundary data 
\begin{eqnarray}
d \vec{A} &=& \overrightarrow{\text{div}} \big(dJ \wedge \Gamma\big) + \overrightarrow{\text{div}} \big( J\, d\Gamma\big) - d\big(\overrightarrow{\langle d J ; \tilde{\Gamma}\rangle }\big),   \label{A-d-bdd} \\
\delta \vec{A} &=& v \label{A-delta-bdd}
\end{eqnarray}    
on $\partial \Omega$, where $v \in W^{m-1,p}(\Omega)$ is a vector valued $0$-form free to be chosen.
\end{Prop}

\Proof
This proposition is a consequence of Lemma \ref{Lemma_basic_smooth} and \ref{Lemma_basic_low-reg}. We summarize the argument here for completeness.  To prove the forward implication and derive \eqref{PDE_A_Cor2}, add  $\delta$ of \eqref{PDE_A_Thm1} to $d$ of the free vector valued function $\delta \vec{A}=v$. This gives \eqref{PDE_A_Cor2}. Restricting \eqref{PDE_A_Thm1} and $\delta \vec{A}=v$ to the boundary gives \eqref{A-d-bdd} - \eqref{A-delta-bdd}.

To prove the backward implication assume first that $m\geq 3$, then take $d$ of \eqref{PDE_A_Cor2} to get 
\begin{eqnarray} \nonumber
\Delta d\vec{A}  &=&  d\delta \Big( \overrightarrow{\text{div}} \big(dJ \wedge \Gamma\big) + \overrightarrow{\text{div}} \big( J\mm d\Gamma\big) - d\big(\overrightarrow{\langle d J ; \tilde{\Gamma}\rangle}\big) \Big)   \cr
&=& d\delta \big( d \big(\overrightarrow{\delta ( J \mm \Gamma )}\big) - d\big(\overrightarrow{\langle d J ; \tilde{\Gamma}\rangle }\big) \big)  \cr
 &=& \Delta \big( d \big(\overrightarrow{\delta ( J \mm \Gamma )}\big) - d\big(\overrightarrow{\langle d J ; \tilde{\Gamma}\rangle }\big) \big) ,
\end{eqnarray}
which is equivalent to
\beq \nonumber
 \Delta w  = 0,
\eeq
with $w$ defined by
\beq \nonumber
w \equiv  d\vec{A} - d \big(\overrightarrow{\delta ( J \mm \Gamma )}\big) + d\big(\overrightarrow{\langle d J ; \tilde{\Gamma}\rangle }\big)  .
\eeq
Thus, since $w$ vanishes on the boundary by \eqref{A-d-bdd}, we conclude that $w=0$ in $\Omega$ which is the sought after equation \eqref{PDE_A_Thm1}. The low regularity case $m=2$ follows by Lemma \ref{Lemma_basic_low-reg}. This completes the proof.
\QED

\appendix

\section{Basic Estimates from Elliptic Regularity Theory}

We review basic elliptic regularity results relevant for the RT-equations \eqref{eqn1} - \eqref{eqn4}. Note, the Laplacian $\Delta=d\delta+\delta d$ acts component-wise on differential forms, so regularity estimates for the scalar Poisson equation extend directly to matrix valued differential forms. We assume from now on that $\Omega$ is a bounded open set in $\R^n$ with smooth boundary, (at least $C^{1,1}$). \\

\noindent {\bf Theorem (Elliptic Regularity):} 
{\it  Let $u \in W^{2,p}(\Omega)$ be a scalar, $\infty > p >1$. Then there exists a constant $C>0$ depending only on $\Omega$, $m,n,p$, such that 
\beq \label{Poissonelliptic_estimate_Lp}
\| u \|_{W^{2,p}(\Omega)} \leq C \Big( \| \Delta u \|_{L^p(\Omega)} + \| u \|_{W^{1,p}(\Omega)} +  \|u \|_{W^{2-\frac{1}{p},p}(\partial\Omega)} \Big).
\eeq } 

Estimate \eqref{Poissonelliptic_estimate_Lp} is equation (2,3,3,1) in \cite{Grisvard}. Estimates for the regularity of the first order equations (\ref{firstorder}) that parallel the estimates for the classical Poisson equation (\ref{Poissonelliptic_estimate_Lp}) are given by the Gaffney inequality, which we now state, (c.f. Theorem 5.21 in \cite{Dac}). 
\\

\noindent {\bf Theorem (Gaffney Inequality):} 
{\it Let $u \in W^{m+1,p}(\Omega)$ be a $k$-form for $m\geq 0$, $p\in (1,\infty)$, $1\leq k\leq n-1$ and (for simplicity) $n\geq2$. Then there exists a constant $C>0$ depending only on $\Omega$, $m,n,p$, such that
\beq \label{Gaffney}
\|u\|_{W^{m+1,p}(\Omega)} \leq C \Big( \|du\|_{W^{m,p}(\Omega)} + \|\delta u\|_{W^{m,p}(\Omega)}+\| u\|_{W^{m+\frac{p-1}{p},p}(\partial\Omega)}\Big). 
\eeq } 

Again, estimate \eqref{Gaffney} for scalar valued differential forms extend to matrix valued differential forms. In this paper, more specifically in Theorem \ref{Thm0}, we only rely on the following elliptic estimate with respect to the $W^{m+1,p}$-norm, for $m\geq 1$, on compactly contained subsets of $\Omega$, which we prove here for completeness.

\begin{Lemma} \label{Lemma_elliptic_estimate}
Let $f \in W^{m-1,p}(\Omega)$ be a  scalar, where $m\geq 1$, $p\in (1,\infty)$. Assume the scalar $u \in W^{m+1,p}(\Omega)$ solves $\Delta u =f$. Then, for any open set $\Omega'$ compactly contained in $\Omega$, there exists a constant $C>0$, depending only on $\Omega'$, $\Omega$, $m,n,p$, such that 
\beq \label{elliptic_estimate_Lp_Lemma}
\| u \|_{W^{m+1,p}(\Omega')} \leq C \big( \| f \|_{W^{m-1,p}(\Omega)} + \| u \|_{W^{m,p}(\Omega)} \big).
\eeq
\end{Lemma}

\Proof
Let $\Omega'$ be an open set that is compactly contained in $\Omega$. Equation (9.36) of Theorem 9.11 in \cite{GilbargTrudinger} gives estimate \eqref{elliptic_estimate_Lp_Lemma} in the case $m=1$, that is,
\beq \label{elliptic_estimate_techeqn1}
\| u \|_{W^{2,p}(\Omega')} \leq C \big( \| f \|_{L^p(\Omega)} + \| u \|_{L^p(\Omega)} \big).
\eeq
Now, from the definition of the $W^{m+1,p}$-norm, we find that
\begin{eqnarray} \label{elliptic_estimate_techeqn2}
\| u \|_{W^{m+1,p}(\Omega')}  \leq \sum_{|\alpha|\leq m-1} \| D^\alpha u \|_{W^{2,p}(\Omega')} ,
\end{eqnarray}
where $\alpha$ denotes a standard multindex and $D^\alpha$ the corresponding combination of partial derivatives, c.f. \cite{Evans}. Differentiating $\Delta u =f$ by $D^\alpha$ and applying  \eqref{elliptic_estimate_techeqn1} to each term on the right hand side of \eqref{elliptic_estimate_techeqn2} then yields 
\begin{eqnarray} \nonumber
\| u \|_{W^{m+1,p}(\Omega')}  
&\leq & C \sum_{|\alpha|\leq m-1}\big( \| D^\alpha f \|_{L^p(\Omega)} + \| D^\alpha u \|_{L^p(\Omega)} \big) \cr
&\leq & C \big( \| f \|_{W^{m-1,p}(\Omega)} + \| u \|_{W^{m-1,p}(\Omega)} \big) \cr
& \leq & C \big( \| f \|_{W^{m-1,p}(\Omega)} + \| u \|_{W^{m,p}(\Omega)} \big) ,
\end{eqnarray}
which is the sought after estimate \eqref{elliptic_estimate_Lp_Lemma}.
\QED

\section*{Conclusion, Discussion and Outlook}\label{Sec_shock_discussion}\label{RTsection}

We have reduced the problem of whether a connection $\Gamma\in W^{m,p}(\Omega)$ can be smoothed one order by coordinate transformation, under the assumption $d\Gamma\in W^{m,p}(\Omega)$, to the problem of finding solutions $(J,\Gammati,A)$ of the RT-equations  (\ref{eqn1})  -(\ref{eqn4}) with boundary data \eqref{bdd1} within the regularity class $J,\Gammati\in W^{m+1,p}(\Omega)$, $A\in W^{m,p}(\Omega)$. The main difficulty for constructing an appropriate existence theory for (\ref{eqn1}) - (\ref{eqn4}) is that the right hand sides are coupled nonlinearly, and \eqref{bdd1} is not standard Dirichlet or Neumann boundary data.   Existence for the case $\Gamma\in W^{m,p}(\Omega)$, for $p>n$, $m\geq1$, is established in authors' companion paper \cite{ReintjesTemple_ell2}.  

The case $\Gamma, d\Gamma$ in $L^{\infty},$ relevant to regularity singularities in GR shock wave theory, is delicate, and is the topic of authors' current research.  In particular, the condition \eqref{J_bdd_Thm1} requires $Curl(J)=0$ on the boundary of the domain, so Lipschitz continuity of $J$ is a regularity too weak to assign boundary conditions in a classical (strong) sense. (The method of assigning Dirichlet data in our companion paper \cite{ReintjesTemple_ell2} is sufficient to resolve this problem, even in the case of $L^\infty$ connections.) Moreover, the existence theory for the linear Poisson equation admits Calderon-Zygmund singularities when the source functions are in $L^{\infty}$, so solutions of the RT-equations can fail to be two levels more regular than the sources.   Note that {\it consistency} of the RT-equations (\ref{eqn1}) - (\ref{eqn4}) is not an issue even in the $L^\infty$ case, because any Lipschitz continuous connection can be transformed to a connection no smoother than $L^\infty$ by application of a $C^{1,1}$ coordinate transformation, and reversing this, the inverse Jacobian together with $\Gammati$ will solve the Riemann-flat condition \eqref{Riemann_flat_J-form} for the transformed connection, where $\Gammati$ is the Lipschitz connection in the original coordinates we started with. 

We can explore the possibility that Calderon-Zygmund singularities might be ruled out by imposing further conditions on $\Gamma$, for example assuming $\Gamma$ lies in the space BMO (Bounded Mean Oscillation), a space containing $L^\infty$, or assuming $\Gamma$ lies in BV (Bounded Variation), a subspace of $L^\infty$ appropriate for shock wave theory, \cite{Smoller,Dafermos}; or, since the problem is local, by modifying $\Gamma$ off an arbitrarily small neighborhood of a given point. We also have the freedom to choose $v$ in system (\ref{eqn1}) - (\ref{eqn4}). 

Consider briefly the freedom to change $\Gamma$ for the problem of regularity singularities. The problem is to establish the existence of a coordinate transformation $x\to y$ that smooths the connection in a neighborhood of any given point $p$.  For this purpose, there is no loss of generality in taking $\Omega$ to be $B_{\epsilon}(p)$, the ball of radius $\epsilon$ centered at $p$ in $\mathbb{R}^n$. Moreover, since the Riemann-flat condition is a point-wise condition, there is no loss of generality in replacing $\Gamma$ by a connection $\Gamma'_{\epsilon}$ which agrees with $\Gamma$ on $B_{\epsilon}(p)$, but extends $\Gamma$ beyond $B_{\epsilon}(p)$ by an auxiliary smooth connection.  To make this precise, let $\Gamma_\infty \in C^\infty(\mathbb{R}^n)$ be such an auxiliary connection and define 
\begin{eqnarray}  \nonumber
\Gamma^*_{\epsilon}=(1-\phi^\epsilon_r)\,\Gamma_{\infty}+\phi^\epsilon_r\,\Gamma, \label{Gammaprime}
\end{eqnarray}
where $\phi^\epsilon_r$ is the standard smooth cutoff function satisfying $\phi^\epsilon_r(x)=1$ if $x\in B_\epsilon(p)$ and  $\phi^\epsilon_r(x)=0$ if $x\in B_r(p)^c$, where $B_r(p)^c$ denotes the complement of $B_r(p)$ in $\mathbb{R}^n$, $r>\epsilon$.  Clearly, $d\Gamma^*_{\epsilon}\in L^\infty(\mathbb{R}^n).$ Thus, if we can solve the RT-equations with $\Gamma^*$ in place of $\Gamma$, we can employ Theorem \ref{Thm_main} to conclude that the Riemann-flat condition holds for the original $\Gamma$, in a neighborhood of $p$.  Note here that we have the freedom to choose $\Gamma_\infty$ and $\Gammati_{\infty}$ to be a known solution of the Riemann-flat condition at the start, and can use $\epsilon$ as a small parameter in an existence theory.    We conclude that there is {\it enormous} freedom, all the freedom to choose $\Gamma_{\infty},v$ and $\epsilon,r$, available to modify the sources in (\ref{eqn1}) - (\ref{eqn4}) in order to avoid Calderon-Zygmund singularities when the sources of the RT-equations are in $L^{\infty}$.  Addressing the problem of regularity singularities for connections of regularity lower than $W^{1,p}$, $p>n$, is the topic of authors current research.\footnote{Since the writing of this paper, we resolved the $L^\infty$ case in \cite{ReintjesTemple_ell4} by obtaining optimal connection regularity $W^{1,p}$, $p<\infty$, but the case $p=\infty$ of Calder\'on Zygmund singularities is still open.}

\section*{Acknowledgements}

The authors thank Jos\'e Natario and the Instituto Superior T\'ecnico for supporting this research, funding two visits of the second author at the institute in Lisbon. The authors thank Craig Evans for suggesting reference \cite{Dac}, and for helpful comments on Calderon-Zygmund theory. We thank John Hunter for discussions and for directing us to reference \cite{Grisvard}, and Jorge da Silva, Pedro Gir\~ao, Steve Shkoller and Kevin Luli for helpful discussions.   Finally, we thank Joel Hass for writing a nice public science article about our research program in the 2018 UC-Davis yearly Mathematics Newsletter.

\providecommand{\bysame}{\leavevmode\hbox to3em{\hrulefill}\thinspace}
\providecommand{\MR}{\relax\ifhmode\unskip\space\fi MR }
\providecommand{\MRhref}[2]{%
  \href{http://www.ams.org/mathscinet-getitem?mr=#1}{#2}
}
\providecommand{\href}[2]{#2}

\end{document}